\documentclass[preprint,12pt,authoryear]{elsarticle}

\usepackage{lmodern}
\usepackage[top=2.5cm,bottom=2.5cm,left=2.5cm,right=2.5cm]{geometry}
\usepackage{ifthen}
\newcommand{\includefig}[2][]{\IfFileExists{#2}{\includegraphics[#1]{#2}}{\fbox{\begin{minipage}{0.9\linewidth}\centering[figure not yet generated]\end{minipage}}}}
\usepackage{amsmath,amssymb,amsthm}
\usepackage{bm}                          %
\usepackage{booktabs}                    %
\usepackage{array}                       %
\usepackage[ruled,linesnumbered]{algorithm2e}  %
\usepackage{lineno}                      %
\usepackage{tikz}                        %
\usetikzlibrary{arrows.meta,calc,bending}
\usepackage[hidelinks,breaklinks=true]{hyperref}
\makeatletter
\newcounter{subsubsubsection}[subsubsection]
\renewcommand\thesubsubsubsection{\thesubsubsection.\@arabic\c@subsubsubsection}
\newcommand\subsubsubsection{\@startsection{subsubsubsection}{4}{\z@}%
   {-2.5ex\@plus -1ex \@minus -.2ex}%
   {1ex \@plus .2ex}%
   {\normalfont\normalsize\bfseries}}
\newcommand*\l@subsubsubsection{\@dottedtocline{4}{7em}{4em}}
\makeatother
\newcommand{\bs}[1]{\boldsymbol{#1}}

\newcommand{\pder}[2]{\dfrac{\partial #1}{\partial #2}}
\newcommand{\ppder}[2]{\dfrac{\partial^2 #1}{\partial #2^2}}
\newcommand{\mixder}[3]{\dfrac{\partial^2 #1}{\partial #2\,\partial #3}}
\newcommand{\Cel}{\mathbf{C}^{e}}
\newcommand{\Cep}{\mathbf{C}^{ep}}
\newcommand{\Idev}{\mathbb{P}_{\mathrm{dev}}}
\newcommand{\sgn}{\operatorname{sgn}}
\newcommand{\tr}{\operatorname{tr}}
\newcommand{\sym}{\operatorname{sym}}
\newcommand{\skw}{\operatorname{skw}}

\journal{}

\begin{document}

\begin{frontmatter}

\title{Buildability Assessment of 3D-Printed Concrete Structures\\
       Using Smoothed Mohr--Coulomb Plasticity with Isotropic Hardening}

\author[bam]{Saif-Ur-Rehman\corref{cor1}}
\ead{saif.ur-rehman@bam.de}
\author[bam]{Annika Robens-Radermacher}
\author[tue]{R.~J.~M. Wolfs}
\author[bam]{J\"{o}rg F. Unger}

\cortext[cor1]{Corresponding author}

\affiliation[bam]{organization={Bundesanstalt f\"{u}r Materialforschung und -pr\"{u}fung (BAM)},
             country={Germany}}
\affiliation[tue]{organization={Department of the Built Environment, Eindhoven University of Technology},
             city={Eindhoven}, country={the Netherlands}}

\begin{abstract}
Extrusion-based 3D concrete printing builds structural
elements layer upon layer without formwork, which leaves the freshly
deposited material to carry the growing weight of the structure while it is
still gaining strength and stiffness. The largest number of layers that can be
printed before the structure fails, its buildability, is therefore a central
quantity in process design. It is governed by the coupled evolution of
geometry, loading and material resistance, and its prediction rests above all
on how accurately the constitutive model represents fresh concrete. Existing
finite element frameworks pair gradual pre-failure hardening only with smooth
yield surfaces, which leaves the Mohr--Coulomb surface that best describes fresh
concrete restricted to perfect plasticity.

This work develops a buildability framework that combines a
smoothed Mohr--Coulomb yield surface with nonlinear isotropic hardening of
the cohesion. The model is formulated in an updated Lagrangian finite element setting and
uses an activation field to represent layer deposition.
Failure is identified from a tangent-stiffness eigenvalue criterion that
agrees with the analytical self-weight buckling load of a slender wall and is
shown to be insensitive to geometric imperfection, unlike the divergence of
the nonlinear solver and other deformation-based criteria in common use. 

The framework is validated against two printed benchmark cases. It reproduces
the measured collapse of a straight wall within one printed layer and predicts
the collapse of a hollow cylinder closer to the experiments than the existing modeling frameworks. A parametric study over six diameters and six hardening rates shows that
buildability peaks at intermediate diameters and rises with the hardening
rate. For each case we also record how much of its strength the material
has used up when the structure fails. Slender cylinders fail with much of that
strength still in reserve, whereas intermediate ones use nearly all of it. The
two failure modes are therefore identified as the limits of one process rather than separate failure
mechanisms.
\end{abstract}


\end{frontmatter}

\section{Introduction}
\label{sec:introduction}

In extrusion-based 3D concrete printing, structural elements are formed
directly from fresh concrete. The material is deposited layer upon layer under
digital control and without the moulds and formwork on which conventional
construction depends. This enables automated construction, expands the range
of geometries that can be produced, and may reduce material use by depositing
concrete according to structural
demand~\citep{Buswell2018,DeSchutter2018,Wangler2016}. However, reductions in
embodied carbon are not guaranteed because printable mixtures often require
high binder contents to achieve extrudability and shape
stability~\citep{Kazemian2017}.

The feature that gives the process its appeal also creates its central
structural difficulty. With no formwork to provide lateral support, the
partially built element carries its own weight from the moment the first layer
is laid. The load at the base grows with every layer while the material there
is still in an early and mechanically weak state. Stiffness and strength
increase over minutes to hours through thixotropic re-flocculation and the onset
of cement hydration, so that both the cohesion and the elastic modulus rise
several-fold within the first hours after
deposition~\citep{Roussel2018,Wolfs2018}. The structural response is therefore
governed by the competition between increasing self-weight and the
age-dependent load-bearing capacity.

When the load exceeds the available resistance, the structure fails during
printing through one of two mechanisms observed experimentally, plastic
collapse and elastic buckling~\citep{Suiker2018,wolfs_suiker2019,Suiker2020}.
Plastic collapse is a strength mechanism, in which the stress at the base of the
element reaches the yield strength of the material. Elastic buckling is a
stability mechanism, in which the element loses geometrical stability while the
stress it carries is still below that strength. Which mechanism governs depends
on geometry, print speed, layer dimensions and material properties. The maximum
number of layers deposited before either failure mode occurs defines the
buildability of the
structure~\citep{Suiker2018,Chang2023}.

Buildability has been assessed experimentally, analytically and numerically, and
all three routes were reviewed in our earlier work~\citep{SaifUrRehman2026}.
Experimental and analytical approaches are summarised briefly here because the
present study focuses on numerical assessment and its constitutive basis. Early
experimental studies determined buildability by depositing successive layers
until deformation of the lower layers became significant~\citep{Le2012}.
Subsequent tests reproduced the stress state during deposition more closely and
related shape stability to mix design~\citep{Kazemian2017,Panda2018,Perrot2016}.
These tests provide direct, material-specific estimates of buildability, but
their results depend on the mixture, test geometry and boundary conditions. The
cost of full-scale printing further limits their use for design exploration.

Analytical models estimate buildability using closed-form criteria. Rheological
descriptions treat the fresh material as a thixotropic yield-stress fluid and
compare its evolving yield stress with the stress carried by the lowest
layer~\citep{Roussel2018,Perrot2016}. This stress-based comparison predicts
plastic collapse only. Buckling requires a separate stability criterion based
on stiffness. Rheological descriptions also apply only while the material still
flows. The transition from fluid-like to solid-like behaviour is gradual, and
both descriptions may be applicable while the material stiffens. At the
timescales considered in structural buildability analyses, the deposited layers
are commonly represented as a solid body. Structural-mechanics models therefore
treat the printed element as a slender member under its own weight.
\citet{Suiker2018} reduced the equilibrium of such a member to a fourth-order
differential equation whose lowest eigenvalue gives the critical buckling
length, and paired that bifurcation analysis with a plastic-collapse limit to
obtain closed-form design graphs with the ageing of the material represented
explicitly. The framework was subsequently validated through comparisons with
finite element simulations and printing
experiments~\citep{wolfs_suiker2019,Suiker2020}. Such models are transparent and
quick to interpret, but they are confined to idealised straight walls and cannot
represent the non-uniform stress states and large deformations associated with
complex printed geometries.

Numerical approaches represent fresh concrete either as a fluid or as a solid.
Fluid-mechanics models resolve the extrusion of individual filaments and the
shape of the first few deposited layers. Computational fluid dynamics with
generalised Newtonian or Bingham laws, discrete particle methods and particle
finite element descriptions of the free surface all reproduce the deposited
geometry with high
fidelity~\citep{Comminal2020,Mollah2021,Wolfs2021,Mechtcherine2014,Rizzieri2024}.
Extending fluid-based simulations from filament deposition over seconds to
structural build-up over multiple layers and several minutes is computationally
demanding~\citep{Rizzieri2026}. Once
the deposited material has stiffened into a quasi-solid, its response is
governed by solid-body mechanics, which is the setting adopted here.

The solid-mechanics route was established by \citet{Wolfs2018}, who simulated
layer-by-layer printing with a Mohr--Coulomb yield criterion embedded in a
rate-independent plasticity model rather than in a viscous flow law. The
elastic and strength parameters were modelled as age-dependent and calibrated
using uniaxial compression and triaxial shear tests~\citep{Wolfs2019}.
Deposition was represented by activating element layers in sequence, and failure
was assumed when any material point first reached the yield surface. The model
captured the buckling of straight walls accurately~\citep{wolfs_suiker2019}, but
for a hollow cylinder it overpredicted the measured buildability by about
60\%. Subsequent studies examined the sources of this discrepancy.

Part of the gap was traced to modelling choices rather than to the material
law. Because the prescribed print path does not follow the deformation of
previously deposited layers, tying each new layer to the deformed surface
introduces artificial strain. Contact formulations avoid this kinematic
constraint and allow the new layer to settle under
gravity~\citep{Ooms2021}. Idealising the rounded filament cross-section as a
rectangle overstates the contact area between layers and therefore the bending
stiffness of the stack. Representing the deposited rather than idealised layer
geometry reduced the predicted buildability by as much as
40\%~\citep{Reinold2024,An2026}. For the cylinder
studied by \citet{LiuSun2021}, accounting for the reduced effective contact
thickness of the rounded layers reduced the predicted buckling height from
approximately 51 layers to the measured 29 layers. The effect is large because
the bending stiffness that resists buckling scales with the cube of that
thickness. These corrections concern the geometry and boundary conditions of
the printed assembly. In contrast, the effect of material resistance is governed by the
constitutive law considered in the present work.

\begin{figure}[tbp]
\centering
\begin{tikzpicture}[>=Latex,xscale=0.95,yscale=1.0]

  \draw[very thick,-{Latex[length=3mm]}] (0,0) -- (0,5.4) node[above left] {Stress};
  \draw[very thick,-{Latex[length=3mm]}] (0,0) -- (10.3,0) node[below] {Strain};

  \draw[thick,domain=0:9.6,samples=120,smooth]
        plot (\x,{4.4*(\x/4.5)*exp(1 - \x/4.5)});

  \draw[dashed,thick] (0,0) -- (1.5,4.4) -- (9.6,4.4);

  \node[anchor=south,align=center] at (3.6,4.85)
        {Elastic--perfectly plastic assumption};
  \draw[->] (1.7,4.92) -- (1.55,4.48);

  \node[anchor=north,align=center] at (5.95,2.95)
        {Stress--strain curve of early\\ age 3D printed concrete};
  \draw[->] (6.5,3.0) -- (6.95,3.90);

\end{tikzpicture}
\caption{Schematic uniaxial compressive stress--strain response of early-age
  3D-printed concrete. The dashed line is the elastic--perfectly plastic
  idealisation, which caps the strength at a constant plateau, whereas the
  solid curve represents a typical gradual hardening response up to peak
  strength followed by post-peak softening. Redrawn after \citet{Liu2022}.}
\label{fig:ss_schematic}
\end{figure}
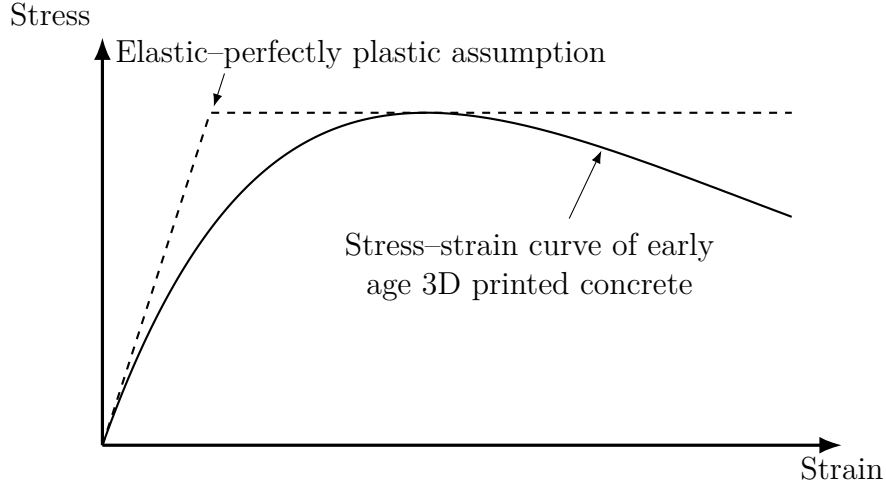

Two constitutive features considered here are pressure dependence and strain
hardening. Fresh concrete is frictional as well as cohesive, so its shear
strength rises with confining pressure, as confirmed by triaxial
tests~\citep{Wolfs2019}. A Mohr--Coulomb or Drucker--Prager yield surface
represents this pressure dependence. Fresh concrete also gains strength
gradually as plastic strain accumulates rather than reaching its full resistance
at first yield. Figure~\ref{fig:ss_schematic} contrasts this measured response
with the elastic and perfectly plastic idealisation, which assigns the peak
strength at first yield and maintains it during subsequent plastic deformation.

\citet{Liu2022} addressed both features by comparing two separate models for
printed cylinders and walls. The Mohr--Coulomb model was kept elastic and
perfectly plastic because they parameterised its cohesion by concrete age
alone. A nonlinear hardening law fitted to uniaxial compression data was
instead incorporated into a Drucker--Prager model, whose smooth cone
avoids the numerical difficulty posed by the hexagonal corners of the
Mohr--Coulomb surface. Hardening in the Drucker--Prager model reduced the error
in the predicted cylinder buildability but did not eliminate the discrepancy
with the experiment.

A parallel effort has enriched the constitutive description with further
inelastic mechanisms. Finite-strain viscoelasticity captures the stiffness
relaxation between depositions~\citep{Nedjar2022}. Hydration kinetics couple
the strength gain to the degree of reaction~\citep{Wang2022}, and ageing damage
models add separate tensile and compressive surfaces~\citep{Chen2025}. Lattice
models reproduce both failure modes using a different numerical
formulation~\citep{Chang2022a,Chang2022}. \citet{An2024} evaluated four
constitutive models for four printed structures and obtained more accurate
predictions with Mohr--Coulomb and concrete damage plasticity than with von
Mises or Drucker--Prager. These results indicate that pressure sensitivity is
important for prediction accuracy. These additional mechanisms improve
physical fidelity but introduce parameters that require dedicated experimental
calibration. Calibrating an ageing damage law with creep and hydration kinetics
demands a campaign of triaxial, creep and curing tests that is rarely feasible
with standard laboratory equipment. The enlarged parameter space also makes it
harder to attribute a predicted change in buildability to a single mechanism.

Previous models have generally represented pressure dependence and nonlinear
hardening separately. The Mohr--Coulomb model of \citet{Wolfs2018} represents
pressure dependence but assumes perfect plasticity, whereas the
Drucker--Prager model of \citet{Liu2022} incorporates nonlinear hardening. Our
earlier von Mises model also incorporates nonlinear hardening but neglects
pressure dependence~\citep{SaifUrRehman2026}. The present formulation combines
pressure dependence and nonlinear cohesion hardening using the smoothed
Mohr--Coulomb surface of \citet{Abbo1995}. A perfectly plastic model calibrated
near peak strength assigns the peak resistance at first yield and may therefore
overpredict buildability. The hardening is governed by one additional parameter,
calibrated from the same uniaxial compression data required by a standard
Mohr--Coulomb model. Damage is omitted because the simulated failures reported
in Section~\ref{sec:sens_eta} occur at or before peak strength. Predicted
buildability is therefore governed by the pre-peak response shown in
Figure~\ref{fig:ss_schematic}.

The definition of failure is the second modelling choice. Buildability is
commonly determined from the layer at which the Newton--Raphson iteration ceases
to converge~\citep{wolfs_suiker2019}, the first material point reaches the yield
surface~\citep{Wolfs2018,Chang2023} or a prescribed deformation limit is
exceeded~\citep{Ooms2021,SaifUrRehman2026}. These criteria either depend on a
prescribed threshold or reflect numerical behaviour that need not coincide with
structural instability. Solver non-convergence may occur after the structure has
entered a stable post-buckling path, while first yield does not necessarily
imply collapse. A prescribed deformation limit depends on the selected
allowable displacement. The first loss of stability is therefore identified
from the first zero crossing of the smallest eigenvalue of the tangent stiffness
matrix. This criterion reproduces the analytical self-weight buckling load and
is insensitive to small geometric imperfections.

The material model and failure criterion are embedded in an updated Lagrangian
formulation for layer-wise growth under self-weight. The contributions of this
work are
\begin{enumerate}
  \item a smoothed Mohr--Coulomb model with nonlinear cohesion hardening for
        buildability assessment of 3D-printed concrete,
  \item a tangent-eigenvalue criterion for the first loss of stability during
        printing,
  \item validation against printed wall and hollow-cylinder experiments, and
  \item a parametric analysis linking geometry, hardening rate, buildability
        and strength mobilisation.
\end{enumerate}

Section~\ref{sec:theory} develops the smoothed Mohr--Coulomb model with
nonlinear hardening and derives the updated Lagrangian formulation and stability
criterion. Section~\ref{sec:results} defines the printing simulation, states the
buildability criterion in the form a growing structure requires, and compares the
candidate stopping criteria on a slender wall whose self-weight buckling height is
known in closed form. Section~\ref{sec:results_all} then validates the framework
against the straight wall and hollow cylinder printed by
\cite{Wolfs2018,Wolfs2019}, and examines the combined effects of cylinder diameter
and hardening rate on buildability and failure mode. A strength-mobilisation
metric is used there to separate instability-dominated from
collapse-dominated failure.
Section~\ref{sec:conclusions} summarises the main findings.
\section{Theoretical Framework}
\label{sec:theory}

Buildability results from the quasi-static response of a growing self-weight structure whose strength rises as it deforms. We model it with small-strain
elastoplasticity on a smoothed Mohr--Coulomb yield surface with nonlinear
cohesion hardening, embedded in an updated Lagrangian formulation with a
co-rotational stress update. The innovation compared to the state-of-the-art models presented in Section~\ref{sec:introduction} is the inclusion of nonlinear hardening of the frictional yield surface, which is made possible by smoothing the corners and the apex of the Mohr--Coulomb hexagon.

The formulation separates the response of the material point from that of the
structure. At the material point, the constitutive law returns the stress, the
history variables and the algorithmic tangent. Equilibrium is then enforced on a
configuration updated at every load step. The material stiffness drives the
equilibrium iteration, and the stiffness contributed by the stress already
present is added to it to decide stability.

The constitutive theory of Sections~\ref{sec:constitutive}
and~\ref{sec:yield_surface} is written in tensor notation with the summation
convention. The algorithmic developments of Section~\ref{sec:return_mapping} and
of \ref{app:tangent} are matrix algebra and are written in the Mandel
representation, in which a symmetric second-order tensor of components $a_{ij}$
is stored as the six-component array
\begin{equation}
  \bigl[a_{11},\,a_{22},\,a_{33},\,
        \sqrt{2}\,a_{12},\,\sqrt{2}\,a_{23},\,\sqrt{2}\,a_{13}\bigr]^{\top}
  \label{eq:mandel}
\end{equation}
and a symmetric fourth-order tensor as a $6\times6$ matrix \citep{Mandel1965}.
The $\sqrt{2}$ factors on the shear entries make the double contraction
$a_{ij}b_{ij}$ the Euclidean inner product of the two arrays, which turns the
local solution of the return mapping into ordinary matrix algebra
\citep{rosenbusch2025}.

\subsection{Thermodynamic Constitutive Setting}
\label{sec:constitutive}

Early-age concrete deforms plastically well before it fails, and its shear resistance rises with that deformation towards a ceiling. We represent this pre-peak gain with a single scalar hardening variable $\alpha$ whose conjugate force raises the cohesion from its initial value $c_0$ towards a saturated value $c_\infty$. Strains are small within a load step, so
the total strain is decomposed additively as
\begin{equation}
  \bs{\varepsilon} = \bs{\varepsilon}^e + \bs{\varepsilon}^p,
  \qquad
  \varepsilon_v^e = \tr\bs{\varepsilon}^e,
  \qquad
  \bs{\varepsilon}^{e\prime}
    = \bs{\varepsilon}^e - \tfrac{1}{3}\varepsilon_v^e\,\bs{1},
  \label{eq:additive_split}
\end{equation}
with $\varepsilon_v^e$ the elastic volume change, $\bs{\varepsilon}^{e\prime}$
the part of the elastic strain that changes shape at constant volume and
$\bs{1}$ the second-order identity tensor. The Helmholtz free energy is the sum
of an isotropic elastic part and a hardening part,
\begin{equation}
  \Psi(\bs{\varepsilon}^e,\alpha)
  = \tfrac{1}{2}\kappa\,(\varepsilon_v^e)^2
  + \mu\,\bs{\varepsilon}^{e\prime}\!:\!\bs{\varepsilon}^{e\prime}
  + (c_\infty-c_0)
    \left(\alpha-\tfrac{1}{\omega}+\tfrac{1}{\omega}e^{-\omega\alpha}\right),
  \label{eq:free_energy}
\end{equation}
where $\kappa=E/[3(1-2\nu)]$ and $\mu=E/[2(1+\nu)]$ are the bulk and shear
moduli. The hardening part is chosen so that its derivative saturates, since the
strength of the fresh material grows towards a finite limit rather than without
bound. The
Coleman--Noll argument gives the stress and the internal force driving the evolution of the hardening variable as the
derivatives of this potential,
\begin{align}
  \bs{\sigma}
  &= \pder{\Psi}{\bs{\varepsilon}^e}
   = \kappa\,\varepsilon_v^e\,\bs{1} + 2\mu\,\bs{\varepsilon}^{e\prime},
  \label{eq:stress_from_energy}\\[4pt]
  \beta
  &= \pder{\Psi}{\alpha}
   = (c_\infty-c_0)\!\left(1-e^{-\omega\alpha}\right).
  \label{eq:beta_from_energy}
\end{align}
The first relation is isotropic linear elasticity,
$\bs{\sigma}=\Cel\!:\!\bs{\varepsilon}^e$. The second is the internal force
conjugate to $\alpha$ and carries the whole of the strength gain. It
vanishes when there is no plastic deformation and approaches $c_\infty-c_0$ as $\alpha$ grows,
at a rate set by $\omega$. Added to the initial cohesion, it gives the current
cohesion of the yield surface of Section~\ref{sec:yield_surface},
\begin{equation}
  c(\alpha) = c_0 + \beta
            = c_0 + (c_\infty-c_0)\!\left[1-e^{-\omega\alpha}\right].
  \label{eq:hardening}
\end{equation}

\subsection{Smoothed Mohr--Coulomb Yield Surface}
\label{sec:yield_surface}

Frictional strength depends on the confining pressure and on the shape of the
deviatoric stress, so the yield surface is written in the stress invariants
\begin{align}
  I_1 &= \sigma_{ii},
  \qquad
  s_{ij} = \sigma_{ij} - \tfrac{1}{3}I_1\,\delta_{ij},
  \label{eq:I1}\\[4pt]
  J_2 &= \tfrac{1}{2}s_{ij}s_{ij},
  \qquad
  J_3 = \det(\bs{s}),
  \qquad
  \theta = \tfrac{1}{3}\arcsin\!\left(
      \frac{-3\sqrt{3}\,J_3}{2\,J_2^{3/2}}\right).
  \label{eq:J2}
\end{align}
Here $I_1$ is three times the mean stress, $J_2$ measures the intensity of the
shear and $\theta\in[-\pi/6,\pi/6]$ is the Lode angle, which locates the stress
state around the deviatoric plane and is what makes the Mohr--Coulomb section a
hexagon rather than a circle.

The hexagonal section is not differentiable at its corners. At
$\theta=\pm\pi/6$, the gradient of the surface is undefined and the return
mapping has no unique flow direction. It is this difficulty that led the closest
earlier work to place its hardening law on a smooth Drucker--Prager cone instead
\citep{Liu2022}.
The Abbo--Sloan construction \citep{Abbo1995} removes it by replacing each
corner with a $C^2$-continuous polynomial blend over a transition angle
$\theta_T$. Away from the corners, the shape function is
\begin{equation}
  K(\theta,\varphi)
  = \cos\theta - \tfrac{1}{\sqrt{3}}\sin\varphi\,\sin\theta
  \qquad\text{for }|\theta|\le\theta_T,
  \label{eq:K_smooth}
\end{equation}
with $\varphi$ the friction angle. Within the corner region, it is replaced by the
blend given in \ref{app:smoothing}. The yield surface is then $F=0$ with
\begin{equation}
  F(\bs{\sigma},\alpha)
  = \frac{I_1}{3}\sin\varphi
    + \sqrt{J_2\,K^2(\theta,\varphi) + a^2\sin^2\!\varphi}
    - \underbrace{\bigl[c_0+\beta(\alpha)\bigr]}_{c(\alpha)}\cos\varphi,
  \label{eq:yield}
\end{equation}
where $a$ is a small stress that rounds the apex so the gradient stays defined
in hydrostatic tension as well. The pressure term $I_1\sin\varphi/3$ carries the frictional strength and grows with confinement. The deviatoric term is the shear intensity measured on the smoothed hexagon through $K(\theta,\varphi)$. The cohesion term $c(\alpha)\cos\varphi$ carries the strength the material has built up, and through \eqref{eq:hardening} it expands the surface isotropically as printing proceeds.

\begin{figure}[!t]
\centering
\includefig[width=\textwidth]{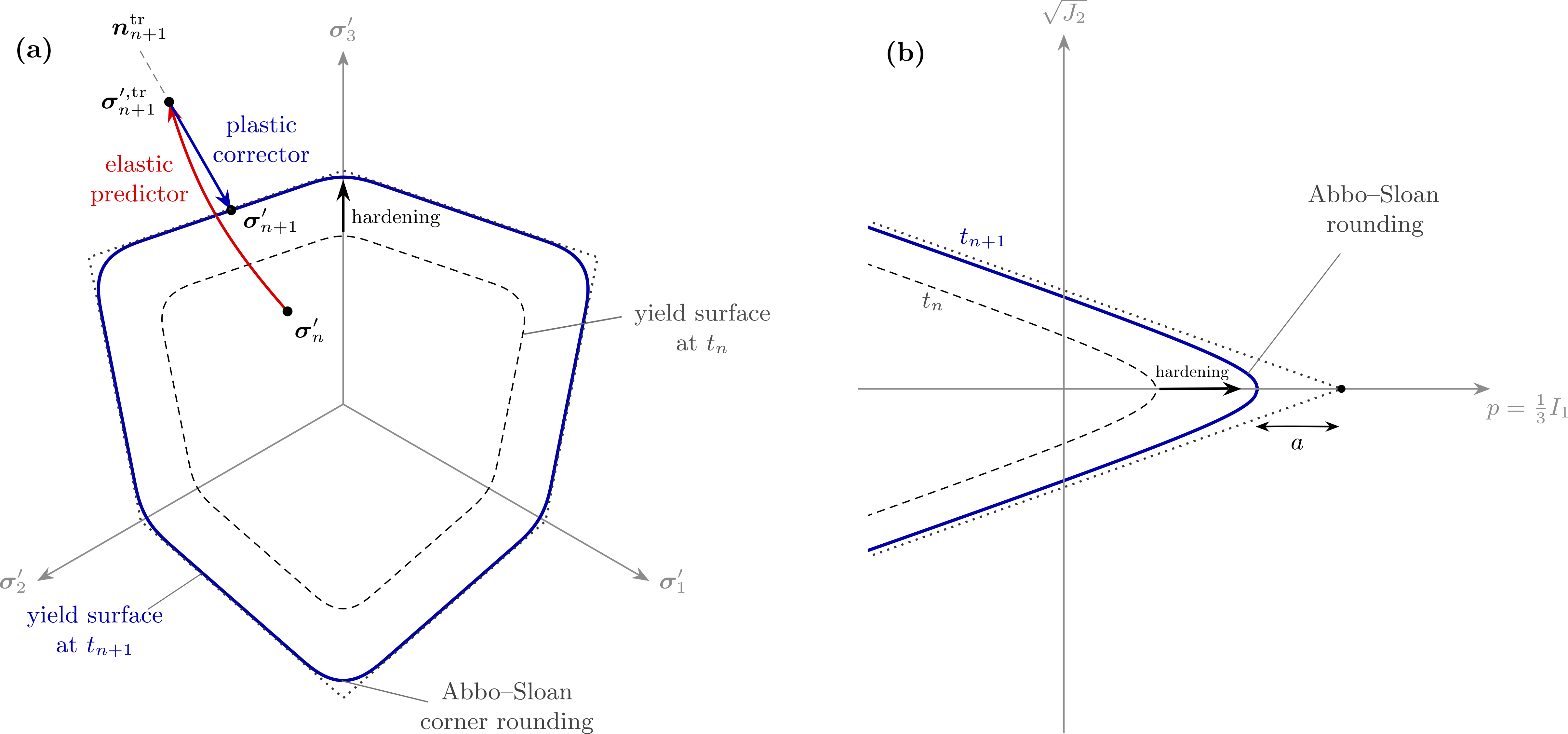}
\caption{Smoothed Mohr--Coulomb yield surface with nonlinear cohesion
hardening. (a)~Deviatoric ($\pi$-plane) section. The nonlinear cohesion
hardening of \eqref{eq:hardening} expands the surface isotropically from its
converged state at $t_n$ to the updated state at $t_{n+1}$, while the
Abbo--Sloan construction of \eqref{eq:K_smooth} rounds the otherwise singular
Mohr--Coulomb corners over the transition angle $\theta_T$. The arrows show the
implicit return mapping of \eqref{eq:r_sigma}--\eqref{eq:r_alpha}. The elastic
predictor carries the deviatoric stress $\bs{\sigma}'_n$ to the trial state
$\bs{\sigma}^{\prime,\mathrm{tr}}_{n+1}$ outside the hardened surface, and the
plastic corrector returns it to $\bs{\sigma}'_{n+1}$ on the surface along the
flow direction $\bs{n}^{\mathrm{tr}}_{n+1}$. (b)~Meridional ($p,\sqrt{J_2}$)
section. The friction angle $\varphi$ sets the slope of the cone and the
hardening translates the apex outward, while the tension cut-off parameter $a$
rounds the apex in hydrostatic tension as in \eqref{eq:yield}.}
\label{fig:yield_surface}
\end{figure}

Figure~\ref{fig:yield_surface} shows the geometry of the yield surface.
Panel~(a) is the section in the deviatoric plane, on which the Abbo--Sloan blend
rounds the corners over $\theta_T$ and the cohesion hardening expands the
surface between two consecutive states. Panel~(b) is the meridional section, on
which the friction angle sets the slope of the cone and $a$ rounds the apex.
Table~\ref{tab:mc_params} lists the parameters of the model.

\begin{table}[tbp]
\centering
\caption{Material parameters of the smoothed Mohr--Coulomb model.}
\label{tab:mc_params}
\begin{tabular}{cll}
\toprule
Symbol & Description & Unit \\
\midrule
$E$         & Young's modulus              & Pa  \\
$\nu$       & Poisson ratio                & --  \\
$\varphi$   & Friction angle               & rad \\
$\psi$      & Dilatancy angle              & rad \\
$c_0$       & Initial cohesion             & Pa  \\
$c_\infty$  & Saturated cohesion           & Pa  \\
$\omega$    & Hardening rate               & --  \\
$a$         & Tension cut-off parameter    & Pa  \\
$\theta_T$  & Abbo--Sloan transition angle & rad \\
\bottomrule
\end{tabular}
\end{table}

For an associated material model, the principle of maximum dissipation makes the
plastic flow normal to the yield surface in the generalised stress space,
\begin{equation}
  \dot{\bs{\varepsilon}}^p = \dot{\lambda}\,\pder{F}{\bs{\sigma}},
  \qquad
  \dot{\alpha} = -\dot{\lambda}\,\pder{F}{\beta}
              = \dot{\lambda}\cos\varphi,
  \label{eq:associated_flow}
\end{equation}
with $\dot{\lambda}$ the plastic multiplier. Normality on a frictional surface
ties the plastic volume change to the friction angle. The formulation therefore
retains this thermodynamic evolution of $\alpha$ but admits a plastic potential
of the same form as $F$ with a dilatancy angle $\psi$ in place of $\varphi$,
\begin{equation}
  G(\bs{\sigma})
  = \frac{I_1}{3}\sin\psi
    + \sqrt{J_2\,K^2(\theta,\psi) + a_g^2\sin^2\!\psi},
  \qquad
  a_g = a\,\frac{\tan\varphi}{\tan\psi},
  \label{eq:potential}
\end{equation}
the scaling of $a_g$ placing the apex of $G$ at the apex of $F$. The flow rule
used in the return mapping is therefore
\begin{equation}
  \dot{\bs{\varepsilon}}^p = \dot{\lambda}\,\bs{m},
  \qquad
  \bs{m} = \pder{G}{\bs{\sigma}},
  \qquad
  \dot{\alpha} = \dot{\lambda}\cos\varphi,
  \label{eq:flow_rule}
\end{equation}
which recovers the associative form of \eqref{eq:associated_flow} when
$\psi=\varphi$. We take $\psi=\varphi$ throughout this work, so the model
applied here is associated and the tangent stays symmetric. Loading and
unloading obey the Karush--Kuhn--Tucker conditions
\begin{equation}
  \dot{\lambda}\ge 0,
  \qquad
  F(\bs{\sigma},\alpha)\le 0,
  \qquad
  \dot{\lambda}\,F(\bs{\sigma},\alpha)=0,
  \label{eq:kkt}
\end{equation}
together with the consistency condition $\dot{F}=0$ whenever the material is
yielding.

\subsection{Return Mapping and Algorithmic Tangent}
\label{sec:return_mapping}

The rate equations are integrated by an implicit backward-Euler update
\citep{deSouzaNeto2008}. Given the strain increment $\Delta\bs{\varepsilon}$ and the converged state
$(\bs{\sigma}_n,\alpha_n)$, the elastic trial stress is
\begin{equation}
  \bs{\sigma}_{n+1}^{\mathrm{tr}}
  = \bs{\sigma}_n + \Cel\,\Delta\bs{\varepsilon},
  \label{eq:trial_stress}
\end{equation}
and the step is elastic when this state lies inside the yield surface. When it lies outside, the updated state
$(\bs{\sigma}_{n+1},\Delta\lambda,\alpha_{n+1})$ is the solution of the residual
system
\begin{align}
  \mathbf{r}_\sigma
  &= \bs{\sigma}_{n+1} - \bs{\sigma}_n
     - \Cel\!\left(\Delta\bs{\varepsilon}
       - \Delta\lambda\,\bs{m}(\bs{\sigma}_{n+1})\right)
     = \mathbf{0},
  \label{eq:r_sigma}\\[4pt]
  r_f &= F\!\left(\bs{\sigma}_{n+1},\alpha_{n+1}\right) = 0,
  \label{eq:r_f}\\[4pt]
  r_\alpha
  &= \alpha_{n+1} - \alpha_n - \Delta\lambda\,\cos\varphi = 0,
  \label{eq:r_alpha}
\end{align}
which impose the flow rule, the consistency condition and the hardening
evolution at the end of the step. We solve them by a local Newton iteration on
the eight unknowns, whose Jacobian is
\begin{equation}
  \mathbf{J} =
  \begin{bmatrix}
    \mathbf{I}+\Delta\lambda\,\Cel\,\pder{\bs{m}}{\bs{\sigma}}
      & \Cel\,\bs{m} & \bs{0} \\[4pt]
    \bs{n}^{\top} & 0 & h_\alpha \\[2pt]
    \bs{0}^{\top} & -\cos\varphi & 1
  \end{bmatrix},
  \label{eq:return_map_linearization}
\end{equation}
with $\mathbf{I}$ the $6\times6$ identity, $\bs{n}=\partial F/\partial\bs{\sigma}$
the yield normal and $h_\alpha=\partial F/\partial\alpha$ the hardening slope.
The component expressions for $\bs{m}$, its Hessian
$\partial\bs{m}/\partial\bs{\sigma}$, $\bs{n}$ and $h_\alpha$ are collected in
\ref{app:tangent}.

The same Jacobian also gives the tangent required by the structural problem.
Perturbing
the converged root with respect to the imposed strain while holding all three
residuals at zero gives the algorithmic tangent
\begin{equation}
  \mathrm{d}\bs{\sigma} = \Cep\,\mathrm{d}\bs{\varepsilon},
  \qquad
  \Cep = \left[\mathbf{J}^{-1}\right]_{\bs{\sigma}\bs{\sigma}}\Cel,
  \label{eq:algorithmic_tangent}
\end{equation}
so the tangent consistent with the discrete update is read from the already
factorised Jacobian rather than assembled separately
\citep[Sec.~3.6]{SimoHughes1998}. An elastic step returns $\Cep=\Cel$.
Algorithm~\ref{alg:mc} summarises the point-wise update, and
Figure~\ref{fig:yield_surface}(a) shows the predictor--corrector split it
performs.

\vspace{\intextsep}
\begin{algorithm}[H]
\SetAlgoLined
\caption{Point-wise smoothed Mohr--Coulomb update at a material point.}
\label{alg:mc}
\KwIn{strain increment $\Delta\bs{\varepsilon}$; previous state
      $\bs{\sigma}_n,\alpha_n$.}
\KwOut{updated state $\bs{\sigma}_{n+1},\alpha_{n+1}$; tangent $\Cep$.}
Form the elastic trial stress~\eqref{eq:trial_stress} and evaluate
  $I_1,\bs{s},J_2,J_3,\theta$\;
\eIf{$F(\bs{\sigma}_\mathrm{trial},\alpha_n)\le 0$}{
  elastic step: $\bs{\sigma}_{n+1}=\bs{\sigma}_\mathrm{trial}$,
  $\alpha_{n+1}=\alpha_n$, $\Cep=\Cel$\;
}{
  initialise $(\bs{\sigma}_{n+1},\Delta\lambda,\alpha_{n+1})
              =(\bs{\sigma}_n,0,\alpha_n)$\;
  \While{residual \eqref{eq:r_sigma}--\eqref{eq:r_alpha} not converged}{
    assemble $\mathbf{J}$~\eqref{eq:return_map_linearization} and take a Newton
    step on $(\bs{\sigma}_{n+1},\Delta\lambda,\alpha_{n+1})$\;
  }
  read $\Cep$ from $\mathbf{J}^{-1}$~\eqref{eq:algorithmic_tangent}\;
}
\end{algorithm}

\subsection{Updated Lagrangian Weak Form}
\label{sec:fem}

At the beginning of a load step, the body occupies $\Omega_n$. Inertia is
negligible at printing speeds, so equilibrium in the
current configuration is
\begin{equation}
  \nabla\cdot\bs{\sigma}+\bs{b}=\bs{0}
  \quad\text{in }\Omega_n,
  \label{eq:equilibrium}
\end{equation}
with $\bs{\sigma}$ the Cauchy stress and $\bs{b}=\rho\bs{g}$ the self-weight per
unit current volume, which is the only load a printing simulation carries. The
displacement is prescribed on $\Gamma_D$ and the traction $\bar{\bs{t}}$ on
$\Gamma_N$.

Deflections accumulate as layers are added and change the geometry that must
carry the next one, so the undeformed body cannot serve as the reference. We
therefore reset the reference to $\Omega_n$ at every step in an updated
Lagrangian description, which is preferred here because the undeformed
configuration loses its physical relevance once the material yields
\citep[Sec.~3.4]{deBorst2012}. The weight of each new layer is applied over
several time steps, so the increment of a single step stays small even though the
accumulated deformation does not, and the incremental displacement $\bs{u}$
measured from $\Omega_n$ is expressed through the infinitesimal strain
\begin{equation}
  \bs{\varepsilon}(\bs{u}) = \sym(\nabla\bs{u}).
  \label{eq:small_strain_increment}
\end{equation}
This is the strain measure work-conjugate to the Cauchy stress on the current
configuration, so each step is a small-strain problem posed on $\Omega_n$.
Equilibrium in weak form is the vanishing of the residual
\begin{equation}
  R(\bs{u};\bs{v})
  = \int_{\Omega_n}\bs{\sigma}:\bs{\varepsilon}(\bs{v})\,\mathrm{d}\Omega
  - \int_{\Gamma_N}\bs{v}\cdot\bar{\bs{t}}\,\mathrm{d}\Gamma
  - \int_{\Omega_n}\bs{v}\cdot\bs{b}\,\mathrm{d}\Omega
  = 0
  \label{eq:weak_residual}
\end{equation}
for every admissible virtual displacement $\bs{v}$.

Interpolating the increment as $\bs{u}=\sum_I N_I\bs{a}_I$ reduces the field
problem to one in the nodal values $\bs{a}$. The strain becomes
$\bs{\varepsilon}=\mathbf{B}\bs{a}$ with $\mathbf{B}$ the small-strain
displacement operator, and the internal force vector is
\begin{equation}
  \bs{f}_{\mathrm{int}}(\bs{a})
  = \int_{\Omega_n}\mathbf{B}^{\top}\bs{\sigma}\,\mathrm{d}\Omega,
  \label{eq:global_residual}
\end{equation}
with $\bs{\sigma}$ the true Cauchy stress returned by the material update at the
integration points. Equilibrium is now the
algebraic statement $\bs{f}_{\mathrm{int}}=\bs{f}_{\mathrm{ext}}$, whose tangent
comes from the constitutive law,
\begin{equation}
  \mathbf{K}_M
  = \int_{\Omega_n}\mathbf{B}^{\top}\,\Cep\,\mathbf{B}\,\mathrm{d}\Omega.
  \label{eq:material_stiffness}
\end{equation}
Because the strain measure is linear in the incremental displacement, the
formulation carries no stiffness from the change of geometry itself. That
contribution governs stability under compression and is reinstated for that
purpose in Section~\ref{sec:buckling}.

Large deformation is accounted for instead by two mechanisms. First, the mesh is
advanced once a step has converged,
\begin{equation}
  \bs{x}_{n+1}=\bs{x}_n+\bs{u},
  \label{eq:mesh_update}
\end{equation}
so that all gradients, volume measures and surface normals in the next step are
evaluated on the deformed body. Second, the stress is transported objectively
across the step.

A stress stored in a fixed frame would rotate spuriously as the body tilts, so
the increment is split into a stretch and a spin,
\begin{equation}
  \Delta\bs{L}=\nabla\bs{u},
  \qquad
  \Delta\bs{D}=\sym(\Delta\bs{L}),
  \qquad
  \Delta\bs{W}=\skw(\Delta\bs{L}),
  \label{eq:strain_spin}
\end{equation}
of which the symmetric part is the strain increment $\Delta\bs{\varepsilon}$
handed to the material update. The gradient is formed on the midpoint
configuration
\begin{equation}
  \bs{x}^\mathrm{mid} = \bs{x}_n + \tfrac{1}{2}\bs{u},
  \label{eq:midpoint_config}
\end{equation}
which makes the increment second-order accurate in the step, after which the
body is restored to $\bs{x}_n$ for assembly. The spin defines the
Hughes--Winget rotation \citep{HughesWinget1980}
\begin{equation}
  \mathbf{Q} = \left(\bs{1}-\tfrac{1}{2}\Delta\bs{W}\right)^{-1}
               \left(\bs{1}+\tfrac{1}{2}\Delta\bs{W}\right),
  \label{eq:hughes_winget}
\end{equation}
whose half-step part $\mathbf{Q}^{1/2}=\exp\!\left(\tfrac{1}{2}\ln\mathbf{Q}\right)$
carries the old stress into the midpoint frame,
\begin{equation}
  \tilde{\bs{\sigma}}_n
  = \mathbf{Q}^{1/2,\top}\,\bs{\sigma}_n\,\mathbf{Q}^{1/2},
  \label{eq:stress_rotation}
\end{equation}
and the returned stress back out of it,
\begin{equation}
  \bs{\sigma}_{n+1}
  = \mathbf{Q}^{1/2}\,\tilde{\bs{\sigma}}_{n+1}\,\mathbf{Q}^{1/2,\top}.
  \label{eq:stress_rotation_back}
\end{equation}
Together, these realise the Jaumann rate integrated at the midpoint. The material
update of Section~\ref{sec:return_mapping} therefore acts entirely in the
midpoint frame, and its residuals
\eqref{eq:r_sigma}--\eqref{eq:r_alpha} are satisfied there.

At each equilibrium iteration, the correction $\Delta\bs{a}$ solves
\begin{equation}
  \mathbf{K}_M\,\Delta\bs{a}
  = -\bigl(\bs{f}_{\mathrm{int}}-\bs{f}_{\mathrm{ext}}\bigr),
  \label{eq:newton_linear_system}
\end{equation}
and the step is accepted once the residual norm falls below a tolerance. The
iteration matrix omits the linearisation of the stress transport
\eqref{eq:stress_rotation}, so it is not the exact Jacobian of the discrete
residual and the asymptotic convergence is linear rather than quadratic. The
omitted term is of the order of the step increment, which the layer-wise
schedule keeps small. The penalty is therefore a slower approach to the
tolerance rather than a loss of robustness. Algorithm~\ref{alg:ul} summarises
the incremental solution.

\vspace{\intextsep}
\begin{algorithm}[H]
\SetAlgoLined
\caption{Incremental updated Lagrangian solution with co-rotation.}
\label{alg:ul}
\KwIn{mesh; boundary conditions; load steps $\{t_1,\ldots,t_N\}$.}
\KwOut{displacement and stress fields at all steps.}
\For{each load step $n=0,1,\ldots,N-1$}{
  apply the load increment and set $\bs{a}=\bs{0}$\;
  \For{each equilibrium iteration}{
    form $\nabla\bs{u}$ on the midpoint
    configuration~\eqref{eq:midpoint_config}\;
    transport $\bs{\sigma}_n$ into the midpoint frame
    through~\eqref{eq:hughes_winget}--\eqref{eq:stress_rotation}\;
    update the material at every integration point
    (Algorithm~\ref{alg:mc})\;
    assemble $\bs{f}_{\mathrm{int}}$ and $\mathbf{K}_M$, solve
    \eqref{eq:newton_linear_system} and update $\bs{a}$\;
    \If{converged}{break\;}
  }
  transport the converged stress out of the midpoint frame
  through~\eqref{eq:stress_rotation_back}\;
  accept the step and advance the mesh through~\eqref{eq:mesh_update}\;
}
\end{algorithm}

\subsection{Linearised Tangent and Stability}
\label{sec:buckling}

The material stiffness \eqref{eq:material_stiffness} measures the stress
increment that a strain produces, but not the work that the stress already
present does against a change of geometry. That work is what destabilises a
slender structure under compression, so it has to be recovered before stability
can be judged. Its origin is the quadratic term
\begin{equation}
  \bs{\eta}(\bs{u}) = \tfrac{1}{2}\nabla\bs{u}^{\top}\nabla\bs{u},
  \label{eq:quadratic_strain}
\end{equation}
dropped from the incremental strain \eqref{eq:small_strain_increment}, which
contributes the internal virtual work
$\int_{\Omega_n}\bs{\sigma}:\delta\bs{\eta}\,\mathrm{d}\Omega$. Linearising that
contribution about the converged state with the stress held fixed gives the
initial-stress stiffness
\begin{equation}
  \mathbf{K}_G
  = \int_{\Omega_n}\mathbf{B}_G^{\top}\,\bs{\Sigma}\,\mathbf{B}_G
    \,\mathrm{d}\Omega,
  \label{eq:geometric_stiffness}
\end{equation}
where $\mathbf{B}_G$ returns the plain displacement gradient and $\bs{\Sigma}$
is the block-diagonal arrangement of the current Cauchy stress, acting once on
the gradient of each
displacement component, so that the product returns
$\sigma_{ij}\,\delta u_{k,i}\,\Delta u_{k,j}$
\citep[Eqs.~3.102 and~3.135]{deBorst2012}. Restoring this single term to an
otherwise small-strain formulation is the approximation on which linear
buckling analysis rests, justified as long as the displacement gradients remain
small up to the critical point \citep[Sec.~3.5]{deBorst2012}. It supplies the
destabilising contribution that a bifurcation criterion needs. The tangent that
decides stability is therefore
\begin{equation}
  \mathbf{K}_T = \mathbf{K}_M + \mathbf{K}_G,
  \label{eq:bk_tangent_decomp}
\end{equation}
assembled on the converged deformed configuration at the end of each accepted
step. Compressive stress makes $\mathbf{K}_G$ negative in bending-type modes,
which is the origin of elastic bifurcation under self-weight. A fully consistent
linearisation would add a further term arising from the stress transport of
\eqref{eq:stress_rotation}. That term is of the order of the step increment and
is omitted here, as it is in the equilibrium iteration.

A discrete system under dead loading is in stable equilibrium when the second
variation of its potential energy is positive for every admissible velocity
field. It reaches a critical state of neutral equilibrium as soon as that
quadratic form vanishes for one such field \citep[Sec.~4.4]{deBorst2012},
\begin{equation}
  \dot{\bs{a}}^{\top}\mathbf{K}_T\,\dot{\bs{a}} > 0
  \quad\text{(stable)},
  \qquad
  \dot{\bs{a}}^{\top}\mathbf{K}_T\,\dot{\bs{a}} = 0
  \quad\text{(critical)}.
  \label{eq:bk_stability_condition}
\end{equation}
Instability is therefore the loss of positive definiteness of the tangent on the
degrees of freedom that belong to printed layers and are free. It is applicable along a printing path on which the active domain, the
self-weight, the material age and the tangent all change from layer to layer.

The associated flow rule of Section~\ref{sec:yield_surface} leaves $\Cep$
symmetric and with it $\mathbf{K}_T$, so the eigenvalues of the restricted
operator are real and its smallest few are obtained with a sparse symmetric
eigensolver. The detector for the loss of stability is the smallest algebraic eigenvalue,
\begin{equation}
  \mathbf{K}_T(t)\,\bs{v} = \mu\,\bs{v},
  \qquad
  \mu_1(t) = \min_i \mu_i(t),
  \label{eq:bk_eigenproblem}
\end{equation}
a positive value marking a stable equilibrium and $\mu_1=0$ the onset of
bifurcation in the corresponding mode $\bs{v}_1$. The hardening law admits no
softening branch, so the material tangent cannot lose positive definiteness. A
sign change of $\mu_1$ is therefore a geometric instability rather than a
material one.

\section{Simulation Setup and Stopping Criterion}
\label{sec:results}
A buildability prediction is only as good as the criterion that decides when the
print has failed. This section defines the printing simulation shared by every
case that follows, then weighs the candidate criteria against each other on a
geometry whose buckling height is known in closed form.

\subsection{Printing Simulation Setup}
\label{sec:setup}
The cases that follow share a single quasi-static printing setup. This
subsection defines the parts common to all of them, the geometry and clamped
base, the layer-by-layer activation of self-weight and stiffness, and the
discretisation. Only the material law and the case-specific dimensions change
from one case to the next.

Deposition is slow enough that inertia is negligible, so the quasi-static
balance of momentum of Section~\ref{sec:fem} holds throughout the print. It is
specialised to a self-weight load that grows as material is added, through an
activation field $\chi(\boldsymbol{x},t)\in[0,1]$ that separates printed from
not-yet-deposited material,
\begin{equation}
  \nabla \cdot \big[\chi(\boldsymbol{x},t)\,\boldsymbol{\sigma}(\boldsymbol{x},t)\big]
  + \chi(\boldsymbol{x},t)\,\rho\,\boldsymbol{g} = \boldsymbol{0}.
  \label{eq:setup_equilibrium}
\end{equation}
The field rises continuously from $\chi=0$ to $\chi=1$ across one deposition
interval, so a layer contributes to self-weight and stiffness only as it is
deposited. Each layer then carries its own age clock, started at the instant of
activation, and the constitutive parameters at a point are evaluated at that
local age. The whole mesh is present from the first step and its elements are
switched on as the print reaches them, rather than being created during the
analysis as in element-birth schemes. Because $\chi$ depends on position and time
alone and not on the displacement, it contributes no term of its own to the
tangent \eqref{eq:bk_tangent_decomp} and enters only as a scaling of the
integrand.

\begin{figure}[tbp]
\centering
{\footnotesize
\begin{tikzpicture}[>=Latex,line join=round,
    dim/.style={<->,>=Latex,thin},
    bc/.style={->,>=Latex,thick}]

  \def\Lx{3.4}   %
  \def\Hz{4.8}   %
  \def\dx{0.35}  %
  \def\dz{0.24}  %

  \draw[thick,fill=black!4] (0,0) -- (\Lx,0) -- (\Lx,\Hz) -- (0,\Hz) -- cycle;          %
  \draw[thick,fill=black!8] (0,\Hz) -- (\Lx,\Hz) -- (\Lx+\dx,\Hz+\dz) -- (\dx,\Hz+\dz) -- cycle; %
  \draw[thick,fill=black!12] (\Lx,0) -- (\Lx+\dx,\dz) -- (\Lx+\dx,\Hz+\dz) -- (\Lx,\Hz) -- cycle; %

  \foreach \k in {1,...,8}{
    \pgfmathsetmacro\zz{\k*\Hz/9}
    \draw[black!45,thin] (0,\zz) -- (\Lx,\zz) -- (\Lx+\dx,\zz+\dz);
  }

  \draw[thick] (0,0) -- (\Lx,0) -- (\Lx+\dx,\dz);
  \foreach \i in {0,0.34,...,3.4}{ \draw[thick] (\i,0) -- (\i-0.26,-0.28); }
  \foreach \t in {0.34,0.67,1.0}{
    \draw[thick] ({\Lx+\t*\dx},{\t*\dz}) -- ({\Lx+\t*\dx-0.26},{\t*\dz-0.28}); }

  \draw[bc] (-0.65,{\Hz/2+0.65}) -- (-0.65,{\Hz/2-0.65}) node[midway,left]{$\rho\boldsymbol{g}$};

  \begin{scope}[shift={(-2.1,0.15)}]
    \draw[bc] (0,0) -- (0.72,0) node[right,inner sep=2pt]{$X^{(1)}$};
    \draw[bc] (0,0) -- ({1.35*\dx},{1.35*\dz})
          node[above right,inner sep=1pt]{$X^{(2)}$};
    \draw[bc] (0,0) -- (0,0.72) node[above,inner sep=2pt]{$X^{(3)}$};
  \end{scope}

  \draw[dim] (0,-0.6) -- (\Lx,-0.6) node[midway,fill=white,inner sep=1pt]{$L$};
  \draw[dim] (\Lx+\dx+0.4,\dz) -- (\Lx+\dx+0.4,{\Hz+\dz})
        node[midway,fill=white,inner sep=1pt,rotate=90]{$N_\mathrm{layer}h_\mathrm{layer}$};
  \draw[dim] (0,\Hz+0.18) -- (\dx,\Hz+\dz+0.18)
        node[midway,above left,inner sep=1pt]{$w$};
  \draw[dim] (-0.35,8*\Hz/9) -- (-0.35,\Hz)
        node[midway,left,inner sep=1pt]{$h_\mathrm{layer}$};

  \node[anchor=north] at (\Lx/2,-1.3) {(a) straight wall};

  \def\X{9.0}    %
  \def\Hc{4.2}   %
  \def\ro{2.3}   %
  \def\rio{1.75} %
  \def\ryo{0.62} %
  \def\ryi{0.47} %

  \begin{scope}[yshift=0.4 cm]
  \fill[black!8] (\X-\ro,\Hc) -- (\X-\ro,0)
        arc[start angle=180,end angle=360,x radius=\ro,y radius=\ryo]
        -- (\X+\ro,\Hc)
        arc[start angle=0,end angle=-180,x radius=\ro,y radius=\ryo] -- cycle;

  \draw[thick,dashed] (\X-\ro,0) arc[start angle=180,end angle=0,  x radius=\ro,y radius=\ryo];
  \draw[thick]        (\X-\ro,0) arc[start angle=180,end angle=360,x radius=\ro,y radius=\ryo];
  \draw[thick] (\X-\ro,0) -- (\X-\ro,\Hc);
  \draw[thick] (\X+\ro,0) -- (\X+\ro,\Hc);
  \foreach \k in {1,...,8}{
    \pgfmathsetmacro\zz{\k*\Hc/9}
    \draw[black!45,thin] (\X-\ro,\zz) arc[start angle=180,end angle=360,x radius=\ro,y radius=\ryo];
  }
  \draw[thick,fill=black!8] (\X,\Hc) ellipse[x radius=\ro,y radius=\ryo];
  \draw[thick,fill=white]   (\X,\Hc) ellipse[x radius=\rio,y radius=\ryi];

  \pgfmathsetmacro\rmid{(\ro+\rio)/2}
  \draw[->,>=Latex,thin] (\X,\Hc) -- ({\X-\rmid},\Hc) node[midway,above,inner sep=1pt]{$R$};
  \draw[dim] (\X+\rio,\Hc) -- (\X+\ro,\Hc);
  \node[anchor=south west,inner sep=1pt] at ({\X+\ro-0.1},{\Hc+0.04}){$w$};

  \foreach \t in {182,194,...,358}{
    \draw[thick] ({\X+\ro*cos(\t)},{\ryo*sin(\t)})
              -- ({\X+\ro*cos(\t)-0.24},{\ryo*sin(\t)-0.26}); }

  \draw[bc] ({\X-\ro-0.85},{\Hc/2+0.65}) -- ({\X-\ro-0.85},{\Hc/2-0.65})
        node[midway,left]{$\rho\boldsymbol{g}$};

  \draw[dim] ({\X+\ro+0.6},0) -- ({\X+\ro+0.6},{\Hc})
        node[midway,fill=white,inner sep=1pt,rotate=90]{$N_\mathrm{layer}h_\mathrm{layer}$};
  \end{scope}

  \node[anchor=north] at (\X,-1.3) {(b) hollow cylinder};

  \node[align=center] (tf) at (5.3,5.4) {traction-free\\ surfaces};
  \draw[->,>=Latex,thin] (4.55,5.4) -- (\Lx+0.5*\dx,\Hz-0.1);          %
  \draw[->,>=Latex,thin] (6,5.4) -- ({\X-\ro+0.25},{\Hc-0.15});     %

\end{tikzpicture}%
}
\caption{Geometry, boundary conditions and layer build-up of the two
  validation cases. (a) The straight wall of length $L$, width $w$ and height
  $N_\mathrm{layer}h_\mathrm{layer}$, the slender self-weight-dominated limit.
  (b) The hollow cylinder of mean radius $R$ and wall thickness $w$. In both,
  the base is fully clamped and all other surfaces are traction-free, no
  symmetry conditions are imposed, and material is added layer by layer under
  self-weight $\rho\boldsymbol{g}$. The fixed Cartesian frame
  $(X^{(1)},X^{(2)},X^{(3)})$ used throughout is drawn at the left, with
  $X^{(3)}$ vertical.}
\label{fig:setup_geometry}
\end{figure}
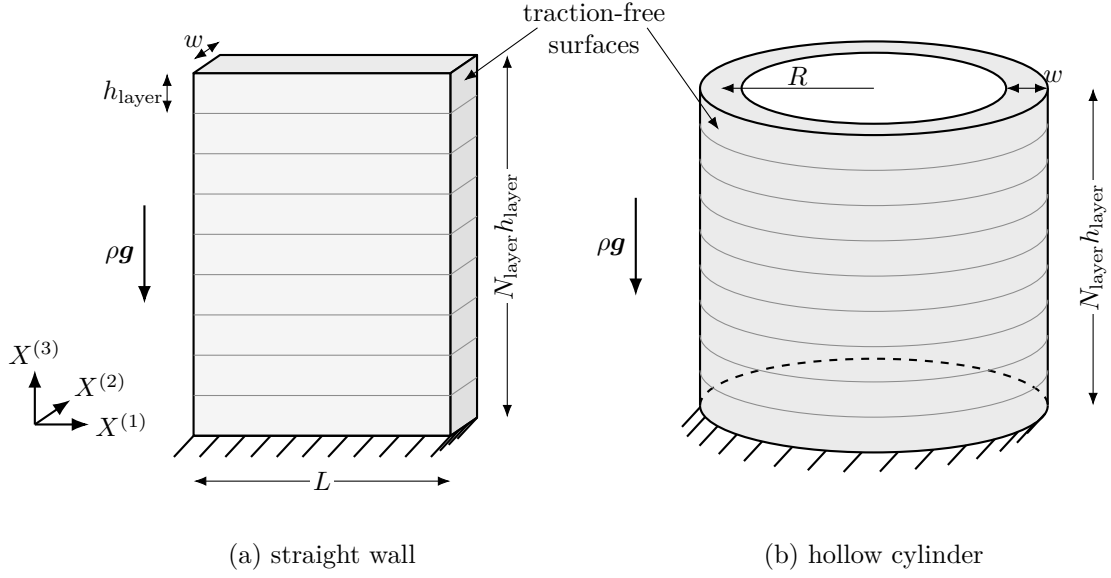

The stopping criteria are evaluated on two geometries, both shown in
Figure~\ref{fig:setup_geometry}. The straight wall has length $L$, width $w$ and
height $N_\mathrm{layer}h_\mathrm{layer}$, and the hollow cylinder has mean
radius $R$ and wall thickness $w$. In both geometries the base
is fully clamped, all other surfaces are traction-free and no symmetry
conditions are imposed, so asymmetric buckling modes remain admissible.

The mesh density and the time step were fixed by a mesh and time-step convergence
study, in which each was refined until the peak von Mises stress and the critical
layer count were insensitive to further refinement. Every result reported below
is computed on the converged discretisation, whose element counts are quoted with
each case.

\subsection{Buildability Criterion}
\label{sec:criterion}

Section~\ref{sec:buckling} states the stability criterion in terms of the tangent
of the discrete system. Applying it to a print requires evaluating it on a
structure whose active domain grows from one step to the next and reading the
instant at which it is first satisfied as a number of layers.

Two eigenvalue indicators are tracked along the build. The detector is that
criterion applied at every accepted step, the sign change of the smallest
eigenvalue $\mu_1$ of the total tangent as the structure actually stands. The
predictor instead poses the classical buckling problem afresh at each accepted
step with the state frozen at the instant $t$,
\begin{equation}
  \bigl(\mathbf{K}_M(t) + \lambda\,\mathbf{K}_G(t)\bigr)\bs{v} = \bs{0},
  \label{eq:bk_classical}
\end{equation}
whose smallest positive root
\begin{equation}
  \lambda_\mathrm{cr}^\mathrm{cl}(t)
  = \min\bigl\{\lambda>0 :
      \det(\mathbf{K}_M(t)+\lambda\,\mathbf{K}_G(t))=0\bigr\}
  \label{eq:bk_lambda_cl}
\end{equation}
is the factor by which the present stress field would have to be scaled to make
the tangent singular. It descends towards unity as compression accumulates, and
its crossing of $\lambda=1$ marks the same event as the detector in the limit of
fine step resolution.

Each indicator marks an instant along the build, which we convert to a layer
count. The detector is triggered between the two consecutive converged steps that
bracket the sign change of $\mu_1$, and linear interpolation places it at
\begin{equation}
  t_\mathrm{cr}
  = t_n
  + \frac{\mu_1(t_n)}{\mu_1(t_n)-\mu_1(t_{n+1})}\,(t_{n+1}-t_n).
  \label{eq:bk_interp}
\end{equation}
Because the printer deposits one layer every $\Delta t_\mathrm{layer}$, any such
time reads directly as a fractional layer count,
\begin{equation}
  N_\mathrm{p}(t) = \frac{t}{\Delta t_\mathrm{layer}},
  \qquad
  N_\mathrm{cr} = N_\mathrm{p}(t_\mathrm{cr}),
  \label{eq:bk_layers_fractional}
\end{equation}
so detector, predictor and collapse share one comparable axis. We report
$N_\mathrm{cr}$ throughout as the buildable layer count, which is the maximum number of
layers printed before failure. It is fractional because a criterion may be met
part-way through the deposition of a layer.

Algorithm~\ref{alg:bk} summarises the evaluation of both eigenvalue indicators
along the build path.

\vspace{\intextsep}
\begin{algorithm}[H]
\SetAlgoLined
\caption{Stability tracking on the converged build path.}
\label{alg:bk}
\KwIn{converged stress $\bs{\sigma}_{n+1}$ and configuration from
      Algorithm~\ref{alg:ul}.}
\KwOut{detector layer $N_\mathrm{cr}$; per-step predictor
       $\lambda_\mathrm{cr}^\mathrm{cl}$.}
\For{each accepted load step}{
  assemble $\mathbf{K}_M$~\eqref{eq:material_stiffness} and
  $\mathbf{K}_G$~\eqref{eq:geometric_stiffness} on the
  converged mesh\;
  restrict $\mathbf{K}_T$ to the printed and free degrees of freedom\;
  compute $\mu_1$~\eqref{eq:bk_eigenproblem} and
  $\lambda_\mathrm{cr}^\mathrm{cl}$~\eqref{eq:bk_lambda_cl}, and record
  $N_\mathrm{p}$~\eqref{eq:bk_layers_fractional}\;
  \If{$\mu_1$ changes sign}{
    interpolate $t_\mathrm{cr}$~\eqref{eq:bk_interp} and store $N_\mathrm{cr}$
    and the mode $\bs{v}_1$\;
  }
}
\end{algorithm}

\subsection{Stopping-Criterion Comparison}
\label{sec:stop2}

Four indicators of the layer at which a printed structure loses load-carrying
capacity are compared:
\begin{enumerate}
  \item the eigenvalue \emph{detector}, the first sign change of the
        smallest tangent eigenvalue $\mu_1(t)$ of $\mathbf{K}_T$
        \eqref{eq:bk_eigenproblem},
  \item the eigenvalue \emph{predictor}, the downward crossing of the
        per-step classical load factor
        $\lambda_\mathrm{cr}^\mathrm{cl}(t)\!\to\!1$
        \eqref{eq:bk_lambda_cl},
  \item the \emph{Newton-divergence} (collapse) layer $N_\mathrm{div}(\xi)$,
        the step at which the nonlinear solve of the equilibrium residual
        \eqref{eq:weak_residual} ceases to converge, and
  \item the \emph{analytical} self-weight buckling height $H_\mathrm{cr}$
        \eqref{eq:bk_h_cr_analytic}, available for the slender wall only.
\end{enumerate}
The aim is to identify which indicator faithfully measures the first loss of
stability. The straight wall serves as the reference case, since its buckling
height is known in closed form and the three numerical indicators can be
checked against it. The cylinder, for which no such closed-form height exists,
then tests whether that agreement carries over. Both cases use a linear-elastic
material whose parameters are held constant in age, so that the only
nonlinearity along the build path is geometric.
Table~\ref{tab:stop2_params} lists the parameters used for the two geometries.

Each case is run for four meshes that share the printing parameters and
material and differ only in a small lateral imperfection of the initial mesh,
\begin{equation}
  \bs{x}^\mathrm{imp}(\bs{X}) = \bs{X} + \xi\,\frac{w}{H}\,X_3\,\bs{e}_2,
  \qquad \xi\in\{0,10^{-4},10^{-3},10^{-2}\},
  \label{eq:bk_imperfection}
\end{equation}
with $\bs{X}$ the reference position and $\bs{e}_2$ the out-of-plane direction,
the first amplitude giving the nominally perfect mesh. An indicator that measures
the first loss of stability of the structure should be essentially independent of
this perturbation, whereas one that measures the failure of a particular
nonlinear continuation should not. The sweep therefore separates the two.

\begin{table}[tbp]
\centering
\caption{Geometry, printing and discretisation parameters for the
         stopping-criterion comparison. Both cases use the same linear-elastic
         material with a modulus that does not age, so that the only
         nonlinearity along the build path is geometric. Element counts are
         given along the length, the width and the layer height for the wall,
         and circumferentially, through the wall and along the layer height for
         the cylinder.}
\label{tab:stop2_params}
\begin{tabular}{llcc}
\toprule
Quantity & Symbol & Straight wall & Cylinder \\
\midrule
Length or mean radius        & $L$, $R$                & $1.0~\mathrm{m}$ & $250~\mathrm{mm}$ \\
Wall width                   & $w$                     & $60~\mathrm{mm}$ & $35~\mathrm{mm}$ \\
Layer height                 & $h_\mathrm{layer}$      & $9.5~\mathrm{mm}$ & $12.5~\mathrm{mm}$ \\
Scheduled layers             & $N_\mathrm{layer}$      & 50 & 50 \\
Density                      & $\rho$                  & $2100~\mathrm{kg\,m^{-3}}$ & $2200~\mathrm{kg\,m^{-3}}$ \\
Layer time                   & $\Delta t_\mathrm{layer}$ & $9.6~\mathrm{s}$ & $6~\mathrm{s}$ \\
Steps per layer              & ---                     & 6 & 6 \\
Young's modulus              & $E$                     & $78~\mathrm{kPa}$ & $78~\mathrm{kPa}$ \\
Poisson ratio                & $\nu$                   & 0.3 & 0.3 \\
Elements                     & ---                     & $(5,4,2)$ & $(48,2,2)$ \\
\bottomrule
\end{tabular}
\end{table}

\subsubsection{Straight wall}
\label{sec:stop2_wall}

The straight wall is idealised as a thin vertical cantilever plate bending about its weak
axis, and the distributed load of its own weight $\rho g$ produces a lateral instability once a critical height is reached. That self-weight buckling height
is \citep{Suiker2018}
\begin{equation}
  H_\mathrm{cr}
  = 1.98635\left(\frac{E\,w^{2}}{12(1-\nu^{2})\rho g}\right)^{1/3}
  = 213.9~\mathrm{mm},
  \label{eq:bk_h_cr_analytic}
\end{equation}
which corresponds to
$N_\mathrm{cr}^\mathrm{an}=H_\mathrm{cr}/h_\mathrm{layer}=22.51$ layers.

Table~\ref{tab:stop2_results} shows that the detector and predictor reproduce this
reference closely. For the perfect mesh and imperfections up to
$\xi=10^{-3}$ they give $N_\mathrm{cr}^\mathrm{det}=22.67$ and
$N_\mathrm{cr}^\mathrm{pred}=22.65$, both within $0.7\%$ of the analytical
value.

The Newton-divergence layer instead lies well above the bifurcation and varies
strongly with $\xi$, falling from $33.17$ layers for the perfect mesh to
$25.50$ for $\xi=10^{-2}$. It therefore measures not the first loss of
stability but how far the chosen imperfect post-buckling path can be continued
before the solve fails.

The largest imperfection also shifts the eigenvalue criteria, to about $23.5$
layers. At $\xi=10^{-2}$, the tilt is no longer a small perturbation. The wall
now carries appreciable bending before bifurcation, so the eigenvalue crossing
occurs on a different equilibrium path from the analytical one
\eqref{eq:bk_h_cr_analytic}. For this wall, the small-imperfection range
$\xi\le10^{-3}$ is thus the clean comparison regime.

\begin{table}[tbp]
\centering
\caption{Stopping-criterion comparison for the linear-elastic wall and cylinder
         of Table~\ref{tab:stop2_params}, across four imperfection amplitudes
         $\xi$. The analytical reference for the wall is
         $N_\mathrm{cr}^\mathrm{an}=22.51$ ($H_\mathrm{cr}=213.9$~mm) of
         \eqref{eq:bk_h_cr_analytic}. No closed form applies to the local mode
         that governs the cylinder.}
\label{tab:stop2_results}
\begin{tabular}{lcccccc}
\toprule
 & \multicolumn{3}{c}{Straight wall} & \multicolumn{3}{c}{Cylinder} \\
\cmidrule(lr){2-4}\cmidrule(lr){5-7}
 & Detector & Predictor & Collapse & Detector & Predictor & Collapse \\
$\xi$ & $N_\mathrm{cr}^\mathrm{det}$ & $N_\mathrm{cr}^\mathrm{pred}$ & $N_\mathrm{div}$
        & $N_\mathrm{cr}^\mathrm{det}$ & $N_\mathrm{cr}^\mathrm{pred}$ & $N_\mathrm{div}$ \\
\midrule
$0$        & 22.67 & 22.65 & 33.17 & 30.67 & 30.59 & 45.50 \\
$10^{-4}$  & 22.67 & 22.65 & 27.50 & 30.67 & 30.59 & 40.67 \\
$10^{-3}$  & 22.67 & 22.65 & 26.67 & 30.67 & 30.59 & 38.33 \\
$10^{-2}$  & 23.50 & 23.49 & 25.50 & 30.50 & 30.36 & 34.17 \\
\midrule
analytical & \multicolumn{3}{c}{$N_\mathrm{cr}^\mathrm{an}=22.51$} & \multicolumn{3}{c}{---} \\
\bottomrule
\end{tabular}
\end{table}

\begin{figure}[tbp]
\centering
\includegraphics[width=\textwidth]{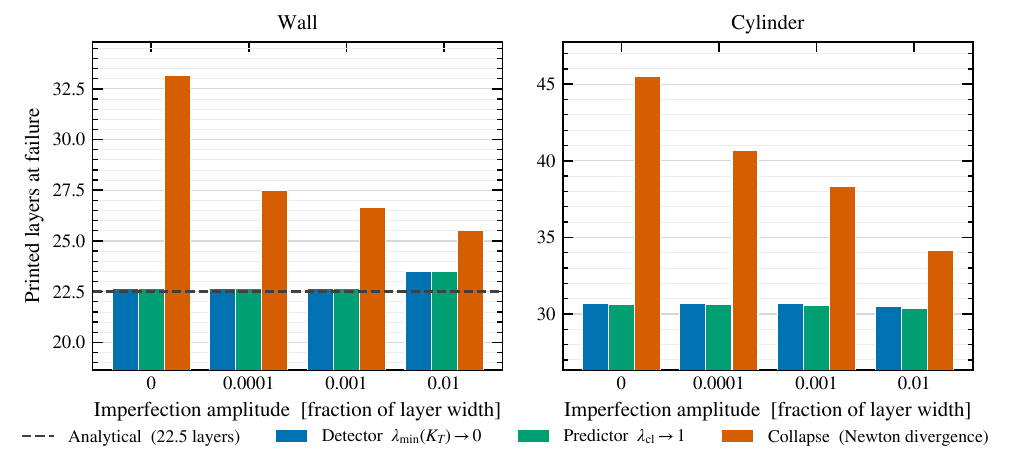}
\caption{Layer at failure versus imperfection amplitude $\xi$ for the
         straight wall (left) and the cylinder (right), comparing the
         eigenvalue detector $\mu_1(K_T)\!\to\!0$, the eigenvalue predictor
         $\lambda_\mathrm{cl}\!\to\!1$, and the Newton-divergence collapse.
         The dashed line in the wall panel is the analytical Greenhill
         reference $N_\mathrm{cr}^\mathrm{an}=22.51$, while no closed form
         applies to the cylinder. Detector and predictor are
         mutually coincident and essentially independent of $\xi$, while
         the collapse layer lies well above the bifurcation and decreases
         with increasing imperfection.}
\label{fig:stop2_bars}
\end{figure}

The left panel of Figure~\ref{fig:stop2_bars} shows this comparison. The
detector and predictor bars sit on the dashed analytical line and barely
move with $\xi$, while the collapse bars stand well above it and drop towards
it as $\xi$ grows.

\subsubsection{Cylinder}
\label{sec:stop2_cyl}

The closed-form self-weight result available for a
hollow cylindrical cantilever describes buckling as a column bending about a
transverse axis \citep{KanahamaSato2022}, which is not the mode that governs
here. The printed shell fails locally in its own wall, for which no closed form
exists, so the comparison rests instead on the mutual agreement of the two
eigenvalue criteria.

Table~\ref{tab:stop2_results} and the right panel of
Figure~\ref{fig:stop2_bars} give the detector
as $N_\mathrm{cr}^\mathrm{det}=30.67$ and the predictor as
$N_\mathrm{cr}^\mathrm{pred}=30.59$ for imperfections up to $\xi=10^{-3}$, a
close agreement to under one tenth of a layer. At the critical layer the two
express the same condition, since a vanishing $\mu_1$ makes
$\mathbf{K}_M+\mathbf{K}_G$ singular and that is exactly
$\lambda_\mathrm{cr}^\mathrm{cl}=1$. Their coincidence is therefore a
consistency check between two differently posed eigenproblems rather than
independent evidence of the instability.

The collapse case again fails later and changes with $\xi$, from $45.50$
layers for the perfect mesh to $34.17$ for $\xi=10^{-2}$, confirming that
Newton divergence overestimates buildability where a stable post-buckling branch
can be traced. By contrast, the detector and predictor barely shift with $\xi$.
Between the perfect mesh and the most imperfect one the detector moves only from
$30.67$ to $30.50$ layers, against a shift of about one layer in the wall.

\begin{figure}[tbp]
\centering
\includegraphics[width=0.85\textwidth]{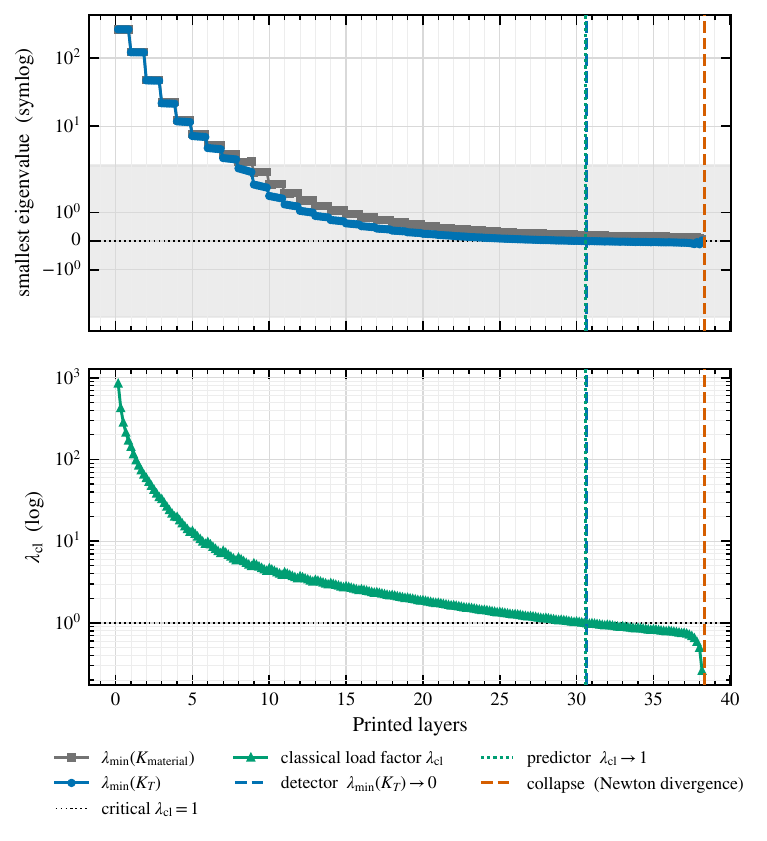}
\caption{Eigenvalue history of the representative cylinder run
         ($\xi=10^{-3}$), against fractional printed layers. \textbf{Top:}
         the smallest material-tangent eigenvalue $\mu_1(K_\mathrm{material})$
         stays positive while the smallest total-tangent eigenvalue
         $\mu_1(K_T)$ crosses zero (shaded band) at the detector layer, so
         the instability is geometric and not material. \textbf{Bottom:} the
         classical load factor $\lambda_\mathrm{cl}$ reaches the critical
         value $1$ at the same layer, confirming the detector. The collapse
         (Newton-divergence) marker lies well beyond both, confirming that
         divergence trails the true bifurcation.}
\label{fig:stop2_eig}
\end{figure}

In Figure~\ref{fig:stop2_eig} the smallest material-tangent
eigenvalue stays positive through the build while the smallest total-tangent
eigenvalue crosses zero, so the instability comes from the geometric stiffness
and not from a loss of material stiffness. The classical load factor reaches unity at the same layer, and
the Newton-divergence marker lies several layers beyond.

\subsubsection{Adopted criterion}
\label{sec:stop2_adopted}

The detector $\mu_1(t)\to0$~\eqref{eq:bk_eigenproblem} is adopted as the
buildability criterion for the rest of the paper. It reproduces the analytical Greenhill bifurcation in the
wall, agrees with the predictor to within a fraction of a layer in the cylinder
where no closed form applies, and is insensitive to small imperfection in
both. The predictor is not used as the governing criterion because its load
factor is meaningful only for a fixed structure under proportional loading,
whereas printing changes the active domain, self-weight, material age and
tangent at every layer, so it is only a frozen-state multiplier that the
detector supersedes once the actual tangent is tracked directly. The
Newton-divergence layer is not carried forward as a criterion, since it
overpredicts buildability in both cases. The simulation is
therefore stopped at the detector crossing, not continued to divergence.

A common alternative defines failure by a threshold out-of-plane deformation, prescribed as a multiple of the layer width \citep{Wolfs2019,Ooms2021} and adopted in the authors' earlier work \citep{SaifUrRehman2026}. The criterion is inexpensive to evaluate and corresponds directly to the deformation observed by a printer operator. However, the prescribed threshold is structure-specific rather than an intrinsic measure of stability, and it is reached only once buckling has grown well past its onset \citep{An2024}. The layer at which it is reached therefore depends on the imposed imperfection, as does the Newton-divergence layer.

\section{Results}
\label{sec:results_all}

The framework of Section~\ref{sec:results} is applied to two printed structures
from the literature, a straight wall and a hollow cylinder. We first validate it
against the collapse measured on each. We then use it to establish how
buildability depends on the printed geometry and on the rate at which the
material gains cohesion, in a sensitivity study.

\subsection{Validation}
\label{sec:validation}

The two validation cases are the straight-wall print of \citet{Wolfs2019} and the
hollow cylinder of \citet{Wolfs2018}. Both references report the collapse measured
in the experiment together with their own finite element prediction of the same
print, so each case places the framework against a measurement and against an
existing numerical study.

\subsubsection{Reference cases and model parameters}
\label{sec:val_params}

Both cases are run with two material laws. The first is the pressure-independent
von Mises model with nonlinear isotropic hardening that the authors used in an
earlier numerical study \citep{SaifUrRehman2026}. It carries a bulk
modulus, a shear modulus, an initial and a saturated yield strength and a
hardening rate, and its yield strength climbs from the initial towards the
saturated value as plastic strain accumulates, in the same exponential manner as
the cohesion law \eqref{eq:hardening}. It is kept here only as a comparator and
is not re-derived. The second is the smoothed Mohr--Coulomb model of
Section~\ref{sec:yield_surface}. Geometry, density, printing parameters and
discretisation are held identical between the two, so any difference in predicted
buildability is attributable to the constitutive law alone.

Parameters are set at deposition and evolve with material age. Ageing acts on the
elastic modulus $E$, the initial and saturated cohesions $c_0$ and $c_\infty$ and
the hardening rate $\omega$ of the Mohr--Coulomb model, and on the bulk modulus,
shear modulus, yield strengths and hardening rate of the von Mises model. Each
grows linearly with material age, at the constant rates listed in
Table~\ref{tab:val_material_params}. The Poisson ratio, friction, dilatancy and
smoothing parameters stay constant. The geometries and printing parameters follow the two references and are listed in
Table~\ref{tab:val_common_params}.

\begin{table}[tbp]
\centering
\caption{Geometry and printing parameters for the two validation cases, each
         common to both material laws. The wall follows \citet{Wolfs2019} and the
         cylinder follows \citet{Wolfs2018}. Element counts are given along
         the length, the width and the layer height for the wall, and
         circumferentially, through the wall and along the layer height for the
         cylinder.}
\label{tab:val_common_params}
\begin{tabular}{llcc}
\toprule
Quantity & Symbol & Straight wall & Hollow cylinder \\
\midrule
Length or centreline radius          & $L$, $R$                  & $1.0~\mathrm{m}$ & $250~\mathrm{mm}$ \\
Wall width                           & $w$                       & $60~\mathrm{mm}$ & $40~\mathrm{mm}$ \\
Layer height                         & $h_\mathrm{layer}$        & $9.5~\mathrm{mm}$ & $10~\mathrm{mm}$ \\
Scheduled layers                     & $N_\mathrm{layer}$        & 30 & 50 \\
Density                              & $\rho$                    & $2100~\mathrm{kg\,m^{-3}}$ & $2070~\mathrm{kg\,m^{-3}}$ \\
Layer time                           & $\Delta t_\mathrm{layer}$ & $9.6~\mathrm{s}$ & $18.85~\mathrm{s}$ \\
Imperfection amplitude               & $\xi$                   & $10^{-3}$ & $10^{-3}$ \\
Elements                             & ---                       & $(5,4,2)$ & $(48,2,1)$ \\
\bottomrule
\end{tabular}
\end{table}

The material parameters are identified from the compression tests reported with
each print. The von Mises parameters are fitted independently at each of four
measured ages, and each of them is then fitted linearly against age to give the
value at deposition and the constant ageing rate quoted in
Table~\ref{tab:val_material_params}. The ageing rates therefore carry the
uncertainty of a four-point regression. We report them as the calibration that
reproduces the measured curves rather than as independently identified material
constants.

\begin{figure}[tbp]
\centering
\includegraphics[width=0.72\textwidth]{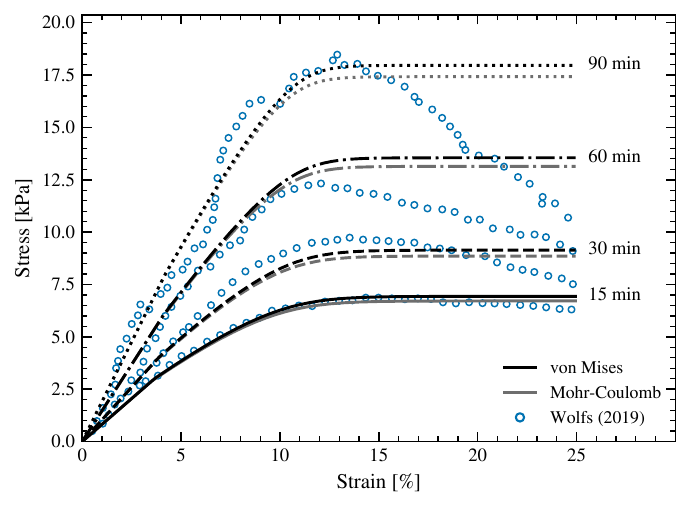}
\caption{Calibrated von Mises and Mohr--Coulomb stress--strain response at four
         material ages against the reference data of \citet{Wolfs2019}. The
         Mohr--Coulomb cohesion is obtained from the von Mises yield strengths
         through the uniaxial conversion of Eq.~\eqref{eq:vm_mc_uniaxial}. The
         Mohr--Coulomb curves fall marginally below the von Mises curves because
         of the Abbo--Sloan smoothing of the yield surface.}
\label{fig:wolfs_stress_strain}
\end{figure}

Both material laws reproduce the reference stress--strain response across the
tested ages (Figure~\ref{fig:wolfs_stress_strain}). The Mohr--Coulomb cohesion is
not fitted independently but mapped from the von Mises strengths. The triaxial
programme of \citet{Wolfs2019} spans three confining pressures, and its branch at
zero confining pressure is an unconfined uniaxial compression test. There the
Mohr--Coulomb condition reduces to a uniaxial relation between yield strength and
cohesion,
\begin{equation}
  \sigma_y = \frac{2c\cos\varphi}{1-\sin\varphi},
  \qquad
  c = \frac{\sigma_y\,(1-\sin\varphi)}{2\cos\varphi}.
  \label{eq:vm_mc_uniaxial}
\end{equation}

The inverse~\eqref{eq:vm_mc_uniaxial} converts each von Mises yield strength, its
ageing rate and the hardening rate into their Mohr--Coulomb counterparts at a
friction angle of $\varphi=20^\circ$. The dilatancy angle is set equal to it,
which leaves the tangent symmetric as Section~\ref{sec:buckling} requires.
Figure~\ref{fig:wolfs_stress_strain} shows that both laws then traverse their
hardening range together in a uniaxial test. Both models thus fit the same
uniaxial compression test through different yield surfaces, and they are reported
together in \ref{app:val_params}.

The Mohr--Coulomb curves sit slightly below the von Mises curves at equal age
because the Abbo--Sloan smoothing rounds the hexagonal cone near its corners,
lowering the yield strength at the sampled states by a few per cent. The
offset is small and conservative and does not affect the conclusions.

\subsubsection{Straight wall}
\label{sec:val_wolfs}

For the straight wall, Table~\ref{tab:wolfs_results} shows that both material laws give the same
buildability, $20.1$ layers. This reproduces the experimental collapse at $21$
layers to within $4.3\%$. The numerical prediction of \citet{Wolfs2019} for the
same wall is $20$ layers, which is within one layer of the present framework prediction. 

For a slender self-weight-dominated wall, buildability is governed by the
global loss of lateral stiffness rather than by the shape of the yield surface.
The Mohr--Coulomb model admits different stress states in the pressure-dependent
lower layers and so reaches a higher peak von Mises stress at failure than the
von Mises model, $5.9~\mathrm{kPa}$ against $4.7~\mathrm{kPa}$, yet the maximum
buildable layer count is unchanged because the instability is a lateral buckling
of the whole wall, not a local exhaustion of strength. The case does not claim
general equivalence of the two yield criteria, only that both reproduce this wall
buckling.

\begin{table}[tbp]
\centering
\small
\caption{Validation results for the Wolfs straight wall. The buildability is
         the detector crossing $\mu_1(K_T)\to 0$ adopted in
         Section~\ref{sec:stop2_adopted}. The last two columns are the printed
         collapse layer and the finite element prediction of \citet{Wolfs2019}.
         The peak von Mises stress is a scalar
         diagnostic and is not the Mohr--Coulomb yield function.}
\label{tab:wolfs_results}
\begin{tabular}{@{}>{\raggedright\arraybackslash}p{0.22\textwidth}
                  >{\raggedright\arraybackslash}p{0.13\textwidth}
                  >{\raggedright\arraybackslash}p{0.15\textwidth}
                  >{\raggedright\arraybackslash}p{0.17\textwidth}
                  >{\raggedright\arraybackslash}p{0.17\textwidth}@{}}
\toprule
Quantity & von Mises & Mohr--Coulomb & Experiment & \citet{Wolfs2019} \\
\midrule
Buildability, $N_\mathrm{cr}^\mathrm{det}$ & 20.1 & 20.1 & 21 & 20 \\
Error against experiment & $-4.3\%$ & $-4.3\%$ & --- & $-4.8\%$ \\
Peak von Mises stress & $4.7~\mathrm{kPa}$ & $5.9~\mathrm{kPa}$ & --- & --- \\
\bottomrule
\end{tabular}
\end{table}

\subsubsection{Hollow cylinder}
\label{sec:val_cylinder}

The second case is the hollow cylinder of \citet{Wolfs2018}, who printed five
nominally identical cylinders until they collapsed and also simulated the print
with their own finite element model.

Table~\ref{tab:cyl_results} shows the results of simulating this cylinder print with
the present approach. The Mohr--Coulomb model reaches the critical state at
$35.9$ layers and the von Mises model at $38.4$. The
Mohr--Coulomb run is repeated with the stiffness and cohesion scaled by
$\pm 17.5\%$ to represent the scatter of the printed material, which brackets the
prediction between $30.8$ and $41.8$ layers. The five prints of
\citet{Wolfs2018} failed at $30$, $25$, $31$, $27$ and $31$ layers, for a mean of
$29$. Their own finite element model predicted a mean of $46$ layers over a range
of $40$ to $53$ that follows from the assumed spread in material properties.

\begin{table}[tbp]
\centering
\small
\caption{Validation results for the hollow cylinder of \citet{Wolfs2018}. The
         buildability is the detector crossing $\mu_1(K_T)\to 0$ adopted in
         Section~\ref{sec:stop2_adopted}. The Mohr--Coulomb range is obtained by
         scaling the stiffness and cohesion by $\pm 17.5\%$ to represent the
         material scatter. The last two columns are the experimental average over
         five prints and the finite element average of \citet{Wolfs2018} with its
         own property-spread range. The peak von Mises stress is a scalar
         diagnostic and is not the Mohr--Coulomb yield function.}
\label{tab:cyl_results}
\begin{tabular}{@{}>{\raggedright\arraybackslash}p{0.22\textwidth}
                  >{\raggedright\arraybackslash}p{0.13\textwidth}
                  >{\raggedright\arraybackslash}p{0.15\textwidth}
                  >{\raggedright\arraybackslash}p{0.17\textwidth}
                  >{\raggedright\arraybackslash}p{0.17\textwidth}@{}}
\toprule
Quantity & von Mises & Mohr--Coulomb & Experiment & \citet{Wolfs2018} \\
\midrule
Buildability, $N_\mathrm{cr}^\mathrm{det}$ & 38.4 & 35.9 (range 30.8--41.8) & 29 (range 25--31) & 46 (range 40--53) \\
Error against experiment & $+32\%$ & $+24\%$ (range $+6\%$ to $+44\%$) & --- & $+59\%$ (range $+38\%$ to $+83\%$) \\
Peak von Mises stress & $7.3~\mathrm{kPa}$ & $9.8~\mathrm{kPa}$ & --- & --- \\
\bottomrule
\end{tabular}
\end{table}

The two models differ by about two and a half layers, with Mohr--Coulomb failing
earlier. This is the effect of pressure dependence that the slender wall did not
show. In the compressed lower layers, confinement from the clamped base raises
the Mohr--Coulomb shear capacity above the shared uniaxial value, so the
Mohr--Coulomb model carries a higher peak von Mises stress at failure,
$9.8~\mathrm{kPa}$ against $7.3~\mathrm{kPa}$. The curved wall also bends, so
parts of the cross-section carry tension. There the pressure-dependent surface is
reached first, and Mohr--Coulomb yields earlier overall. Collapse is still driven
by buckling of the shell, as Section~\ref{sec:stop2} found for the elastic
cylinder, so the difference stays well inside the eleven layers spanned by the
material scatter. Unlike in the wall, however, it is no longer negligible.

Set against the measurements, both predictions lie between the observed collapse
and the finite element prediction of \citet{Wolfs2018}. They over-predict the
experimental mean by about $24\%$ for Mohr--Coulomb and $32\%$ for von Mises,
against $59\%$ for the earlier simulation. The comparison of the two ranges is
sharper still. Our $30.8$ to $41.8$ layers reaches down into the measured spread
of $25$ to $31$, whereas the $40$ to $53$ layers reported by \citet{Wolfs2018}
lies above every one of the five prints. Several effects account for the
remaining gap. We model an idealised geometry and an early-age material state
that the prints never realise exactly, whereas the specimens carried geometric
imperfections and an inhomogeneous material. The as-deposited layer
cross-section also departs from the nominal rectangle assumed here
\citep{Rizzieri2024}. \citet{Wolfs2018} further trace part of their own overshoot
to compressive strengths measured in the laboratory on compacted samples, which
overstate the printed material. A model of the idealised structure should
therefore build higher than the weakest print, and ours stays closer to the
experimental mean than the earlier simulation does.

Across the two benchmarks the framework matches the reference simulation on the
wall, where that simulation already reproduced the print, and reduces the
overshoot on the cylinder from $59\%$ to $24\%$ of the experimental mean.

\subsection{Sensitivity Study}
\label{sec:sensitivity}

Since the cylinder validation showed the greater sensitivity to pressure
dependence, we use that geometry with the Mohr--Coulomb model to study the
effects of hardening and geometry on buildability and failure mode. The cohesion hardening
rate controls strength development with plastic strain, whereas the diameter
governs the structural response to self-weight.

\subsubsection{Parametric design}
\label{sec:sens_design}

The cylinder diameter $D$ is sampled at six values, from a slender
instability-prone column to a wide wall-like shell. The hardening rate
$\omega$ of the cohesion law in Eq.~\eqref{eq:hardening} ranges over six
values, from gradual cohesion hardening over plastic strain to almost immediate
saturation at $c_\infty$.
The full $6\times 6$ grid of cylinder simulations is listed in
Table~\ref{tab:sens_grid}.
The Mohr--Coulomb material parameters are derived from the von Mises parameters
used in our earlier numerical study \citep{SaifUrRehman2026} and do not
correspond to either validation mixture in Section~\ref{sec:validation}. Apart from the hardening rate $\omega$, all material parameters are held constant. The
study therefore examines relative trends rather than the buildable height of a
specific concrete mix. The print time per layer is likewise held at $6$~s for
every diameter, so a given layer carries the same material age across the whole
grid and the comparison isolates the two swept parameters. The deposition speed
this implies grows with the circumference and would not be realisable on every
printer at the largest diameters.

\begin{figure}[tbp]
\centering
\includefig[width=0.72\textwidth]{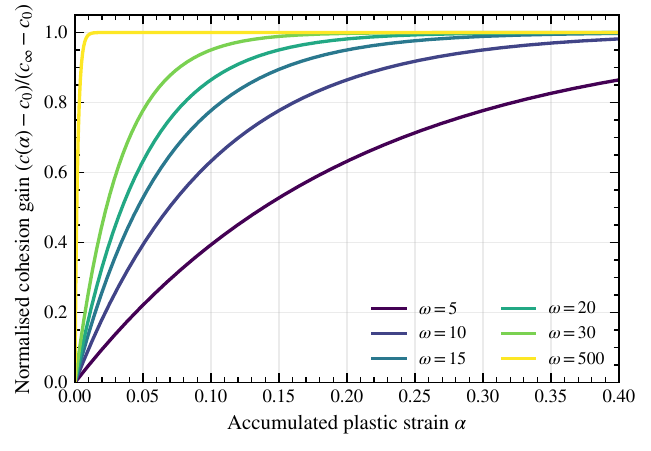}
\caption{Cohesion hardening laws sampled by the study, shown as the
         normalised cohesion gain $(c(\alpha)-c_0)/(c_\infty-c_0)
         =1-e^{-\omega\alpha}$ against accumulated plastic strain $\alpha$.
         Each curve is one value of the hardening rate $\omega$.  Small
         $\omega$ saturates slowly, while $\omega=500$ reaches the saturated
         cohesion almost immediately and approximates perfectly plastic
         behaviour.}
\label{fig:sens_hardening}
\end{figure}

The hardening rate controls how quickly the cohesion climbs from its
initial value $c_0$ towards its saturated value $c_\infty$ as plastic
strain accumulates. Rearranging the hardening law in Eq.~\eqref{eq:hardening}
gives the normalised cohesion gain, which quantifies the fraction of the
increase from $c_0$ to $c_\infty$ attained at the most heavily loaded Gauss
point.
Figure~\ref{fig:sens_hardening} compares the normalised cohesion gain for the
six hardening rates. At the smallest $\omega$, cohesion approaches $c_\infty$
gradually with increasing plastic strain. At $\omega=500$, it reaches
$c_\infty$ after negligible plastic strain, so the subsequent response is
effectively perfectly plastic at the saturated cohesion.

\begin{table}[tbp]
\centering
\caption{Sensitivity-study grid.  Each cell of the $D\times \omega$ matrix
         is one cylinder simulation.  All 36 cells share the same material
         and differ only in the two swept parameters.}
\label{tab:sens_grid}
\begin{tabular}{ll}
\toprule
Quantity & Values \\
\midrule
Cylinder diameter $D$ (mm) & 75, 100, 200, 300, 400, 500 \\
Hardening rate $\omega$       & 5, 10, 15, 20, 30, 500 \\
Layer height (mm)          & 12.5 \\
Layer width  (mm)          & 35   \\
Print time per layer (s)   & 6    \\
\bottomrule
\end{tabular}
\end{table}

\subsubsection{Buildability map}
\label{sec:sens_map}

\begin{figure}[tbp]
\centering
\includefig[width=0.65\textwidth]{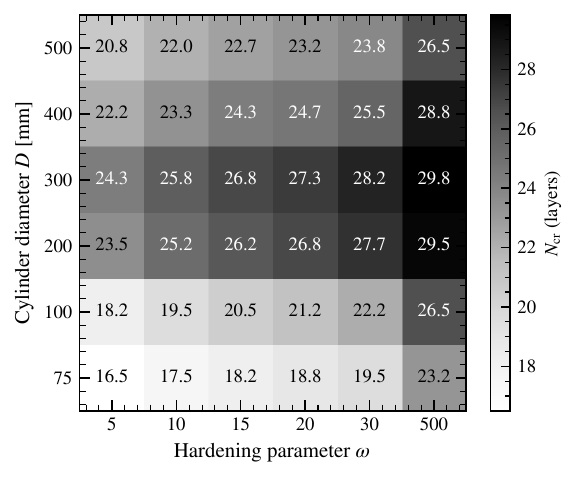}
\caption{Buildability map.  Buildable layer count $N_\mathrm{cr}$ at the
         detector crossing as a function of cylinder diameter $D$ and
         hardening rate $\omega$.  Darker shading marks a higher buildable
         count.  Buildability peaks near $D=300$~mm and increases with
         the hardening rate.}
\label{fig:sens_buildability}
\end{figure}

Figure~\ref{fig:sens_buildability} shows that $N_\mathrm{cr}$ increases
monotonically with $\omega$ for every diameter. Between the lowest and highest
hardening rates, the increase is about 46\% at $D=100$~mm but 23\% at
$D=300$~mm. At fixed hardening rate it is non-monotonic in diameter, rising from
the slender cylinders to a maximum near $D=300$~mm and falling again towards the
widest. The slender columns are limited by global buckling and the widest shells
by buckling of their thin wall, while the intermediate diameters are the most
stable and build the highest. The influence of the hardening rate is strongest
for the slender, instability-prone geometries and weakest for the intermediate
ones, an asymmetry that Section~\ref{sec:sens_eta} explains.

\subsubsection{Observed failure modes}
\label{sec:sens_modes}

The buildable layer count alone does not identify the failure mechanism. Similar
values of $N_\mathrm{cr}$ can result from loss of stability before substantial
hardening or from plastic collapse after extensive hardening. We therefore
examine both the deformed shapes and the material state at failure.

\begin{figure}[tbp]
\centering
\includefig[width=0.92\textwidth]{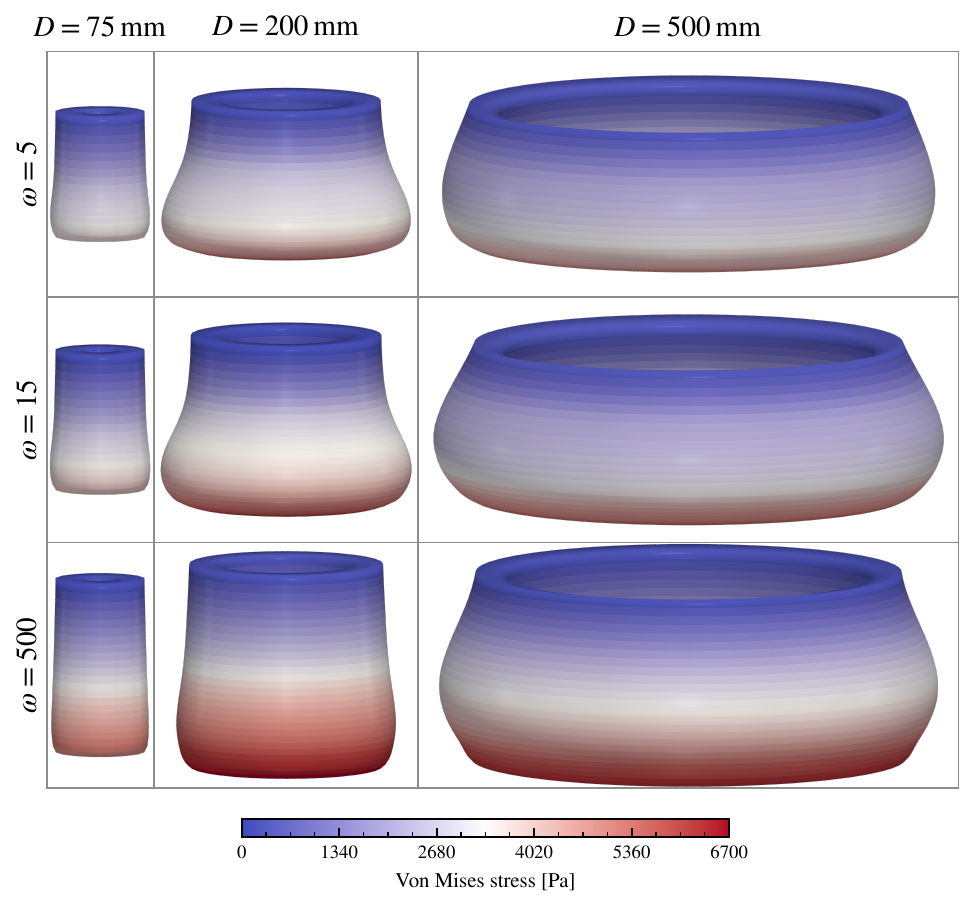}
\caption{Representative deformed shapes at the detected onset of
         failure, coloured by von Mises stress, with the displacements
         magnified five times.  Only the layers deposited at failure are
         shown, and all panels share a common length scale so that the
         true relative sizes of the cylinders are preserved.  Columns are
         diameters $D=75$, $200$ and $500$~mm and rows are hardening
         rates $\omega=5$, $15$ and $500$.  The slender cylinders lean
         about their base, the intermediate cylinders bulge in the lowest
         layers, and the widest cylinders deform into a barrel-like
         profile.}
\label{fig:sens_snapshot}
\end{figure}

Figure~\ref{fig:sens_snapshot} shows representative deformed shapes at the
detected onset of failure, with the displacements magnified five times so that
the small pre-buckling movements are visible. Because the criterion is met at the
bifurcation point, every cylinder is captured before its failure motion has
grown, and all of them are still close to upright. The slender cylinders are
beginning to lean about their base in a single global mode, which is the motion
that carries them over once it develops. The intermediate cylinders bulge
outward in the lowest layers, where the accumulated weight is largest, while the
upper layers stay comparatively straight. The widest cylinders deform over a
large part of their height into a barrel-like profile, each portion of the wall
behaving like a slender plate. The von Mises stress fields support this
interpretation. Stresses remain low in the slender cylinders, reach their
highest values at intermediate diameter, and concentrate in a narrow ring at
the base of the widest cylinders.

The deformed shapes suggest distinct failure mechanisms but cannot establish
whether instability or material yielding initiates failure. Lateral deformation
increases the strain and can drive yielding irrespective of the initial trigger.
A comparison across cylinders failing at different heights and loads therefore
requires a normalised measure of the material state at failure.

\subsubsection{Strength mobilisation at failure}
\label{sec:sens_eta}

The absolute stress at failure is not comparable between cases because cohesion
evolves with both age and plastic strain. The von Mises stress fields in
Figure~\ref{fig:sens_snapshot} illustrate this variation. Across the full grid,
the maximum stress ranges from about $4.9$ to $12.1$~kPa and varies
non-monotonically with diameter, so it does not provide a common measure of
material mobilisation.

The hardening law already supplies the normalisation required. Rearranging
Eq.~\eqref{eq:hardening}, the fraction of the available cohesion gain that a
material point has realised is
\begin{equation}
  \eta = \frac{c(\alpha)-c_0}{c_\infty-c_0} = 1-e^{-\omega\alpha},
  \label{eq:eta}
\end{equation}
which is the normalised cohesion gain already plotted in
Figure~\ref{fig:sens_hardening}. Evaluated at the most heavily loaded
integration point at the instant of failure, it is the strength mobilisation.
It is zero while the point carries its initial cohesion $c_0$ and one when the
cohesion has reached $c_\infty$.

\begin{figure}[tbp]
\centering
\includefig[width=0.72\textwidth]{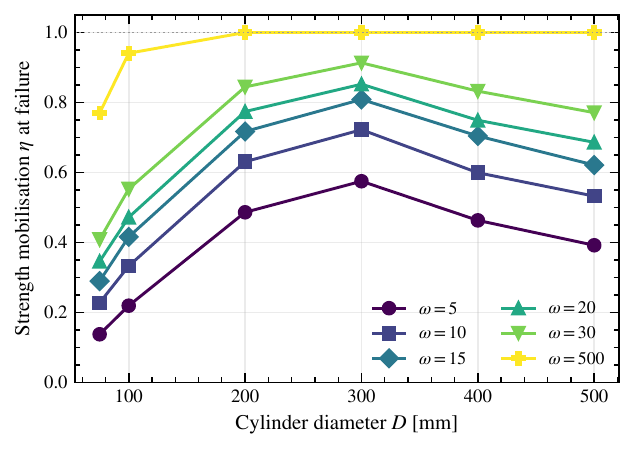}
\caption{Strength mobilisation $\eta$ at failure as a function of
         cylinder diameter $D$, one curve per hardening rate $\omega$.  The
         inverted-U shape peaks at $D=300$~mm and deepens as the
         hardening rate decreases.  The dotted line marks full
         mobilisation $\eta=1$.}
\label{fig:sens_eta_D}
\end{figure}

Figure~\ref{fig:sens_eta_D} reports $\eta$ at failure against diameter, one
curve per hardening rate. The mobilisation follows an inverted-U shape that
peaks at $D=300$~mm, the same diameter at which buildability peaks, and
flattens into a plateau at the perfectly plastic limit, where every cylinder
with $D\ge200$~mm fails fully mobilised. Under the weakest hardening the slender
$D=75$~mm cylinder fails at a mobilisation of only $0.14$ while the intermediate
$D=300$~mm cylinder reaches $0.58$. At fixed diameter the mobilisation rises
monotonically with the hardening rate, and even the slender cylinder reaches
$0.77$ at the perfectly plastic limit. Low mobilisation therefore results from
the combination of slender geometry and weak hardening.
In these cases, structural instability occurs before substantial cohesion
hardening, indicating instability-controlled failure.
Intermediate geometries and the fastest hardening cases instead fail only once
the cohesion is essentially exhausted, the signature of plastic collapse.
Plastic yielding occurs in all cases, but structural instability governs
failure at low mobilisation, whereas plastic collapse governs it near full
mobilisation.

This reading also explains the asymmetry noted in the buildability map. For the
slender geometries that fail by instability at low
mobilisation, a faster hardening rate raises the strength reached before the
instability develops, so the hardening reserve acts as a practical buildability
control. For the intermediate geometries that already fail near full
mobilisation, little reserve remains to be recovered and the hardening rate has
little further effect. The mobilisation therefore gives a material-based reading
of failure that identifies the mechanism and explains the buildability trend at
once, without a separate elastic buckling category, since no case fails before
its most heavily loaded point reaches the initial yield surface.

\section{Conclusions}
\label{sec:conclusions}

This work develops a buildability framework for 3D concrete printing in which
the material response is both pressure-dependent and nonlinearly hardening. A
smoothed Mohr--Coulomb yield surface with nonlinear isotropic hardening of the
cohesion is embedded in an updated Lagrangian finite element formulation with
Jaumann stress rate. A layer activation field grows the structure and its
self-weight together. Stability is read from the total tangent stiffness, which
combines the material stiffness with the geometric stiffness generated by the
stress the structure already carries.

Three definitions of failure were compared on two geometries. In a slender wall
the smallest tangent eigenvalue reaches zero at $22.67$ layers. This is within
$0.7\%$ of the analytical self-weight buckling height, and the classical load
factor gives the same layer. In a cylinder whose local wall mode has no closed
form the two agree to within a tenth of a layer. Both are unaffected by the size
of the imposed geometric imperfection. Newton divergence instead occurs up to
fifteen layers later and drops steadily as the imperfection grows, since it
records how far a post-buckling path can be continued rather than when stability
is lost. The eigenvalue crossing is therefore the most robust of the three, and
we use it as the buildability criterion throughout.

The framework is validated against two printed benchmarks. For the straight wall
of \citet{Wolfs2019} it reproduces the observed collapse at $21$ layers to
within $4.3\%$. Buildability there is set by the global loss of lateral
stiffness, so the von Mises and Mohr--Coulomb surfaces both give $20.1$ layers.
The second benchmark is the hollow cylinder of \citet{Wolfs2018}, where the
cylinder wall is bent and parts of the section carry tension. There the
pressure-dependent surface is reached first, so the Mohr--Coulomb simulation
fails two and a half layers earlier than the von Mises one. The choice of yield
surface therefore shifts the layer count. The Mohr--Coulomb model
predicts collapse at $35.9$ layers against an experimental mean of $29$ over
five prints. Repeating the analysis with upper and lower bound material
properties widens that prediction to between $30.8$ and $41.8$ layers, which
reaches into the observed spread of $25$ to $31$.

A parametric study over six diameters and six hardening rates then examines how
geometry and the rate of cohesion gain together influence buildability. The buildable
layer count rises monotonically with the hardening rate across the sampled range. Against diameter it is
non-monotonic and peaks near $D=300$~mm. The slender cylinders buckle globally
and the widest buckle locally in their thin wall, which leaves the intermediate
diameters the most stable. To account for that asymmetry we introduce the
strength mobilisation at failure, the fraction of the available cohesion gain
that the most heavily loaded point has realised when the structure becomes
unstable, zero at the initial cohesion and one at the saturated cohesion. Under the weakest
hardening the slender $D=75$~mm cylinder fails at a mobilisation of $0.14$ while
the intermediate $D=300$~mm cylinder reaches $0.58$. At the perfectly plastic
limit every cylinder of $D\ge200$~mm fails fully mobilised. Low mobilisation
marks instability reached before the material has spent its strength and full
mobilisation marks plastic collapse. Every failure is nonetheless plastic, since
the most heavily loaded point is already yielding. The two limits differ only in
whether the instability or the exhaustion of strength arrives first.

The cohesion left unspent at failure decides whether a faster hardening rate
helps. A slender geometry becomes unstable with most of its cohesion gain still
unrealised, so a faster gain raises the strength reached before the instability
develops and adds layers. An intermediate geometry already fails fully
mobilised, so a faster gain has almost no effect on the layer count. The
hardening rate is therefore an effective buildability control where failure is
instability-driven, whereas the saturated cohesion sets the limit where it is
collapse-driven.

Two extensions of the material description would strengthen the framework
further. The first is a probabilistic treatment of the calibrated parameters.
Scaling the stiffness and cohesion of the cylinder moves its buildable layer
count by eleven layers, so identifying distributions for those parameters and
propagating them through the
simulation would report buildability as a distribution of attainable layers
rather than a single value. The second is a damage formulation. Isotropic
hardening lets the material gain strength but never lose it, so a softening
branch would carry the framework past the onset of instability and into the
localised cracking and progressive degradation that follow it. The present
formulation already predicts buildability from a physically grounded
constitutive model and an objective stability criterion, and it reproduces the
collapse of two printed structures.

\appendix

\section{Consistent Tangent of the Smoothed Mohr--Coulomb Model}
\label{app:tangent}

This appendix records the gradients, the Jacobian, and the consistent tangent
of the return mapping of Section~\ref{sec:return_mapping}. They are required to
reproduce the constitutive update but not to follow its mechanics.

\subsection{Gradient of the plastic potential}

With the scalar root $r_G=\sqrt{J_2\,K_g^2(\theta,\psi)+a_g^2\sin^2\!\psi}$ and
$K_g=K(\theta,\psi)$, the derivatives of $G$ with respect to the invariants are
\begin{align}
  \pder{G}{I_1} &= \frac{\sin\psi}{3},
  \label{eq:dG_dI1}\\[4pt]
  \pder{G}{J_2} &= \frac{K_g}{2\,r_G}
                   \left[K_g - \tan(3\theta)\,\frac{dK_g}{d\theta}\right],
  \label{eq:dG_dJ2}\\[4pt]
  \pder{G}{J_3} &= \frac{J_2\,K_g\,\tan(3\theta)}{3\,J_3\,r_G}
                   \,\frac{dK_g}{d\theta},
  \label{eq:dG_dJ3}
\end{align}
and the chain rule through the invariants gives
\begin{equation}
  \pder{G}{\bs{\sigma}}
  = \pder{G}{I_1}\,\bs{\delta}
  + \pder{G}{J_2}\,\mathbf{s}
  + \pder{G}{J_3}\,\pder{J_3}{\bs{\sigma}},
  \qquad
  \pder{J_3}{\bs{\sigma}} = \Idev\,\mathbf{s}^2 .
  \label{eq:dG_dsigma}
\end{equation}
The gradient of the yield surface has the same structure with $\varphi$,
$K=K(\theta,\varphi)$ and $r_F=\sqrt{J_2K^2+a^2\sin^2\!\varphi}$,
\begin{align}
  \pder{F}{I_1} &= \frac{\sin\varphi}{3},
  \label{eq:dF_dI1}\\[4pt]
  \pder{F}{J_2} &= \frac{K}{2\,r_F}
                   \left[K - \tan(3\theta)\,\frac{dK}{d\theta}\right],
  \label{eq:dF_dJ2}\\[4pt]
  \pder{F}{J_3} &= \frac{J_2\,K\,\tan(3\theta)}{3\,J_3\,r_F}
                   \,\frac{dK}{d\theta},
  \label{eq:dF_dJ3}\\[4pt]
  \pder{F}{\bs{\sigma}}
  &= \pder{F}{I_1}\,\bs{\delta}
   + \pder{F}{J_2}\,\mathbf{s}
   + \pder{F}{J_3}\,\pder{J_3}{\bs{\sigma}}.
  \label{eq:dF_dsigma}
\end{align}

\subsection{Hessian of the plastic potential}

The second derivative of $J_3$ follows from
\begin{equation}
  T_{ikmn} = \tfrac{1}{2}
    \bigl(\delta_{im}\,s_{nk} + \delta_{in}\,s_{mk}
         + s_{im}\,\delta_{kn} + s_{in}\,\delta_{km}\bigr),
  \label{eq:T4}
\end{equation}
which represents $\partial(\mathbf{s}^2)_{ik}/\partial s_{mn}$ and, mapped to
Mandel form as $\mathbf{T}_\mathrm{mandel}(\mathbf{s})$, gives
\begin{equation}
  \ppder{J_3}{\bs{\sigma}}
  = \Idev\;\mathbf{T}_\mathrm{mandel}(\mathbf{s})\;\Idev .
  \label{eq:d2J3}
\end{equation}
The scalar second derivatives of $G$ are
\begin{align}
  \pder{G}{\theta}
  &= \frac{K_g\,J_2}{r_G}\,\frac{dK_g}{d\theta},
  \label{eq:dG_dtheta}\\[6pt]
  \ppder{G}{\theta}
  &= \frac{J_2}{r_G}
     \left[\left(\frac{dK_g}{d\theta}\right)^{\!2}
           \!\!\left(1-\frac{J_2\,K_g^2}{r_G^2}\right)
           + K_g\,\frac{d^2K_g}{d\theta^2}\right],
  \label{eq:d2G_dtheta2}\\[6pt]
  \mixder{G}{\theta}{J_2}
  &= \frac{K_g}{r_G}\,\frac{dK_g}{d\theta}
     \left(1 - \frac{J_2\,K_g^2}{2\,r_G^2}\right),
  \label{eq:d2G_dthetadJ2}\\[6pt]
  \ppder{G}{J_2}
  &= -\frac{K_g^4}{4\,r_G^3}
     + \frac{(\partial G/\partial\theta)\tan(3\theta)}{2\,J_2^2}
     - \frac{\tan(3\theta)}{2\,J_2}
       \!\left[2\,\mixder{G}{\theta}{J_2}
               - \frac{\tan(3\theta)}{2J_2}\ppder{G}{\theta}
               - \frac{3}{2J_2\cos^2(3\theta)}\pder{G}{\theta}\right],
  \label{eq:d2G_dJ22}\\[6pt]
  \ppder{G}{J_3}
  &= -\frac{\tan(3\theta)}{3\,J_3^2}\pder{G}{\theta}
     + \frac{\tan(3\theta)}{3\,J_3}
       \left[\frac{\ppder{G}{\theta}\,\tan(3\theta)}{3\,J_3}
             + \frac{\partial G/\partial\theta}{J_3\cos^2(3\theta)}\right],
  \label{eq:d2G_dJ32}\\[6pt]
  \mixder{G}{J_2}{J_3}
  &= \mixder{G}{\theta}{J_2}\,\frac{\tan(3\theta)}{3\,J_3}
     - \frac{\tan(3\theta)}{2\,J_2}
       \left[\frac{\ppder{G}{\theta}\,\tan(3\theta)}{3\,J_3}
             + \frac{\partial G/\partial\theta}{J_3\cos^2(3\theta)}\right],
  \label{eq:d2G_dJ2dJ3}
\end{align}
and assembling all contributions by the chain rule gives the full Hessian
\begin{align}
  \ppder{G}{\bs{\sigma}}
  &= \pder{G}{J_2}\,\Idev
   + \pder{G}{J_3}\,\ppder{J_3}{\bs{\sigma}}
   + \ppder{G}{J_2}\!\left(\mathbf{s}\otimes\mathbf{s}\right)
   + \ppder{G}{J_3}\!\left(\pder{J_3}{\bs{\sigma}}\otimes\pder{J_3}{\bs{\sigma}}\right)
  \nonumber\\[4pt]
  &\quad
   + \mixder{G}{J_2}{J_3}\!\left(
       \pder{J_3}{\bs{\sigma}}\otimes\mathbf{s}
       + \mathbf{s}\otimes\pder{J_3}{\bs{\sigma}}\right).
  \label{eq:d2G_dsigma2}
\end{align}

\subsection{Jacobian and consistent tangent}

The return-mapping residual~\eqref{eq:r_sigma}--\eqref{eq:r_alpha} is solved by
Newton iteration on the state vector
$\mathbf{y}=[\bs{\sigma}_{n+1},\Delta\lambda,\alpha_{n+1}]^{\top}$. With the flow
direction $\bs{m}=\partial G/\partial\bs{\sigma}$~\eqref{eq:dG_dsigma}, its
Hessian
$\partial\bs{m}/\partial\bs{\sigma}=\partial^2 G/\partial\bs{\sigma}^2$~\eqref{eq:d2G_dsigma2},
the yield normal $\bs{n}=\partial F/\partial\bs{\sigma}$~\eqref{eq:dF_dsigma}, and
the hardening slope
$h_\alpha=\partial F/\partial\alpha=-\omega(c_\infty-c_0)\,e^{-\omega\alpha}\cos\varphi$,
the $8\times8$ Jacobian of \eqref{eq:return_map_linearization} reads
\begin{equation}
  \mathbf{J}
  = \begin{bmatrix}
      \mathbf{I}+\Delta\lambda\,\Cel\,\pder{\bs{m}}{\bs{\sigma}}
        & \Cel\,\bs{m} & \bs{0} \\[6pt]
      \bs{n}^{\top} & 0 & h_\alpha \\[2pt]
      \bs{0}^{\top} & -\cos\varphi & 1
    \end{bmatrix},
  \label{eq:jacobian}
\end{equation}
updated through
\begin{equation}
  \mathbf{y}^{k+1} = \mathbf{y}^k - \mathbf{J}^{-1}\,\mathbf{r}(\mathbf{y}^k).
  \label{eq:newton}
\end{equation}
On convergence the consistent algorithmic tangent is the top-left block of the
inverse Jacobian scaled by the elastic stiffness,
\begin{equation}
  \Cep = \bigl[\mathbf{J}^{-1}\bigr]_{\bs{\sigma}\bs{\sigma}}\,\Cel,
  \label{eq:Cep}
\end{equation}
so that $\mathrm{d}\bs{\sigma}=\Cep\,\mathrm{d}\bs{\varepsilon}$, with
$\Cep=\Cel$ in the elastic case.

\section{Abbo--Sloan Vertex Smoothing Coefficients}
\label{app:smoothing}

In the vertex region $|\theta|>\theta_T$ the Mohr--Coulomb shape function of
Section~\ref{sec:yield_surface} is replaced by the quadratic blend
\begin{equation}
  K(\theta,\tilde{\varphi})
  = A + B\sin(3\theta) + C\sin^2(3\theta),
  \label{eq:K_vertex}
\end{equation}
where $\tilde{\varphi}$ is $\varphi$ for the yield surface or $\psi$ for the
plastic potential. The coefficients enforce $C^2$ continuity at $\theta_T$
\citep{Abbo1995},
\begin{align}
  c_1 &= \cos\theta_T - \tfrac{1}{\sqrt{3}}\sin\tilde{\varphi}\,\sin\theta_T,
  \label{eq:c1}\\[4pt]
  c_2 &= \sgn(\theta)\,\sin\theta_T
         + \tfrac{1}{\sqrt{3}}\sin\tilde{\varphi}\,\cos\theta_T,
  \label{eq:c2}\\[4pt]
  c_3 &= 18\cos^3(3\theta_T),
  \label{eq:c3}\\[6pt]
  C   &= \frac{-\cos(3\theta_T)\,c_1
               - 3\,\sgn(\theta)\,\sin(3\theta_T)\,c_2}{c_3},
  \label{eq:C_coeff}\\[6pt]
  B   &= \frac{\sgn(\theta)\,\sin(6\theta_T)\,c_1
               - 6\cos(6\theta_T)\,c_2}{c_3},
  \label{eq:B_coeff}\\[6pt]
  A   &= \cos\theta_T
         - \tfrac{1}{\sqrt{3}}\sin\tilde{\varphi}\,\sgn(\theta)\,\sin\theta_T
         - B\,\sgn(\theta)\,\sin(3\theta_T)
         - C\sin^2(3\theta_T).
  \label{eq:A_coeff}
\end{align}
The first two derivatives of $K$ with respect to the Lode angle, required by
the gradients and Hessian of \ref{app:tangent}, are listed in
Table~\ref{tab:dK}.

\begin{table}[tbp]
\centering
\caption{Derivatives of the smoothing function $K(\theta,\tilde{\varphi})$.}
\label{tab:dK}
\begin{tabular}{ccc}
\toprule
Region & $\dfrac{dK}{d\theta}$ & $\dfrac{d^2K}{d\theta^2}$ \\[8pt]
\midrule
$|\theta|\leq\theta_T$ &
  $-\sin\theta - \dfrac{\sin\tilde{\varphi}}{\sqrt{3}}\cos\theta$ &
  $-\cos\theta + \dfrac{\sin\tilde{\varphi}}{\sqrt{3}}\sin\theta$ \\[12pt]
$|\theta|>\theta_T$ &
  $3B\cos(3\theta)+3C\sin(6\theta)$ &
  $-9B\sin(3\theta)+18C\cos(6\theta)$ \\[6pt]
\bottomrule
\end{tabular}
\end{table}

\section{Calibrated Material Parameters for the Validation Cases}
\label{app:val_params}

The parameters obtained by the calibration of Section~\ref{sec:val_params} are
collected in Table~\ref{tab:val_material_params}.

\begin{table}[tbp]
\centering
\small
\caption{Material parameters for both validation cases and both material laws.
         Ageing rates are denoted $A_\bullet$ and carry the units implied by the
         corresponding parameter. The von Mises model is parameterised by the
         bulk and shear moduli and the Mohr--Coulomb model by Young's modulus and
         the Poisson ratio. The two pairs are equivalent and each is quoted in
         the form its own model uses. The Mohr--Coulomb cohesion is the
         uniaxial-compression image of the von Mises yield strength at
         $\varphi=20^\circ$ through \eqref{eq:vm_mc_uniaxial}, so the two laws
         share a common uniaxial calibration and differ only in pressure
         dependence.}
\label{tab:val_material_params}
\begin{tabular}{p{0.12\textwidth}p{0.15\textwidth}p{0.30\textwidth}p{0.30\textwidth}}
\toprule
Model & Parameter group & Straight wall & Hollow cylinder \\
\midrule
von Mises &
Elasticity &
$K=50.593~\mathrm{kPa}$, $\mu=23.351~\mathrm{kPa}$ &
$K=65.0~\mathrm{kPa}$, $\mu=30.0~\mathrm{kPa}$ \\
von Mises &
Hardening strengths &
$\sigma_{y,0}=2.001~\mathrm{kPa}$,
$\sigma_{y,\infty}=4.728~\mathrm{kPa}$, $\omega=64.322$ &
$\sigma_{y,0}=3.00~\mathrm{kPa}$,
$\sigma_{y,\infty}=6.50~\mathrm{kPa}$, $\omega=30$ \\
von Mises &
Ageing rates &
$A_K=20.025~\mathrm{Pa\,s^{-1}}$,
$A_\mu=9.242~\mathrm{Pa\,s^{-1}}$,
$A_{\sigma_{y,0}}=1.110~\mathrm{Pa\,s^{-1}}$,
$A_{\sigma_{y,\infty}}=2.450~\mathrm{Pa\,s^{-1}}$,
$A_{\omega}=1.1702{\times}10^{-2}~\mathrm{s^{-1}}$ &
$A_K=16.667~\mathrm{Pa\,s^{-1}}$,
$A_\mu=7.692~\mathrm{Pa\,s^{-1}}$,
$A_{\sigma_{y,0}}=1.8~\mathrm{Pa\,s^{-1}}$,
$A_{\sigma_{y,\infty}}=2.2~\mathrm{Pa\,s^{-1}}$,
$A_{\omega}=4.5{\times}10^{-3}~\mathrm{s^{-1}}$ \\
\midrule
Mohr--Coulomb &
Elasticity &
$E=60.712~\mathrm{kPa}$, $\nu=0.3$ &
$E=78.0~\mathrm{kPa}$, $\nu=0.3$ \\
Mohr--Coulomb &
Cohesion hardening &
$c_0=700.5~\mathrm{Pa}$,
$c_\infty=1655.5~\mathrm{Pa}$, $\omega=22.519$ &
$c_0=1050.3~\mathrm{Pa}$,
$c_\infty=2275.7~\mathrm{Pa}$, $\omega=10.50$ \\
Mohr--Coulomb &
Friction and smoothing &
$\varphi=\psi=20^\circ$, $\theta_T=26^\circ$, $a=481.13$ &
$\varphi=\psi=20^\circ$, $\theta_T=26^\circ$, $a=721.4$ \\
Mohr--Coulomb &
Ageing rates &
$A_E=24.030~\mathrm{Pa\,s^{-1}}$,
$A_{c_0}=0.389~\mathrm{Pa\,s^{-1}}$,
$A_{c_\infty}=0.858~\mathrm{Pa\,s^{-1}}$,
$A_{\omega}=4.1208{\times}10^{-3}~\mathrm{s^{-1}}$ &
$A_E=20.0~\mathrm{Pa\,s^{-1}}$,
$A_{c_0}=0.630~\mathrm{Pa\,s^{-1}}$,
$A_{c_\infty}=0.770~\mathrm{Pa\,s^{-1}}$,
$A_{\omega}=1.575{\times}10^{-3}~\mathrm{s^{-1}}$ \\
\bottomrule
\end{tabular}
\end{table}

\bibliographystyle{elsarticle-harv}
\bibliography{references}

\end{document}